\documentstyle[12pt]{article}

\newcommand{\sect}[1]{\setcounter{equation}{0}\section{#1}}

\def\bseq{\begin{subequation}}  
\def\eseq{\end{subequation}}
\def\bsea{\begin{subeqnarray}}  
\def\esea{\end{subeqnarray}}

\def\beq{\begin{equation}}
\def\eeq{\end{equation}}
\def\bea{\begin{eqnarray}}
\def\eea{\end{eqnarray}}
\def\bq{\begin{quote}}
\def\eq{\end{quote}}

\newcommand{\EQ}{\begin{equation}}
\newcommand{\EN}{\end{equation}}
\newcommand{\ena}{\end{eqnarray}}

\renewcommand{\a}{\alpha}
\renewcommand{\b}{\beta}

\renewcommand{\d}{\delta}
\newcommand{\th}{\theta}

\newcommand{\pa}{\partial}
\newcommand{\g}{\gamma}

\newcommand{\D}{\Delta}
\newcommand{\e}{\epsilon}

\newcommand{\m}{\mu}

\newcommand{\r}{\rho}

\newcommand{\s}{\sigma}

\newcommand{\reff}[1]{eq.(\ref{#1})}

\newcommand{\shalf}{\frac{1}{2}}
\newcommand{\bl}[1]{\Bigl(#1\Bigr)}
\newcommand{\blt}[1]{\Bigl( #1\Bigr)_\newh}
\newcommand{\cbl}[1]{\Bigl\{ #1\Bigr\}_\newh}
\def\newh{H}
\def\mt{\widetilde{m}}

\def\sh{\mbox{sh}}
\def\ch{\mbox{ch}}

\def\ta{\mbox{$a_{2n-1}^{(2)}$}}
\def\tb{\mbox{$b_n^{(1)}$}}
\def\tc{\mbox{$c_n^{(1)}$}}
\def\td{\mbox{$d_n^{(2)}$}}
\def\tA{\mbox{$A^{(4)}(0,2n)$}}
\def\tB{\mbox{$B^{(1)}(0,n)$}}
\def\tg{\mbox{$g^{(1)}_2$}}

\newcommand{\NP}[1]{Nucl.\ Phys.\ {\bf #1}}
\newcommand{\PL}[1]{Phys.\ Lett.\ {\bf #1}}

\newcommand{\CMP}[1]{Comm.\ Math.\ Phys.\ {\bf #1}}

\renewcommand{\thefootnote}{\fnsymbol{footnote}}

\begin{document}

\newpage
\begin{titlepage}
\begin{flushright}
{CERN-TH.6337/91}\\
{IFUM 413/FT}\\
{hepth@xxx/9201067}
\end{flushright}
\vspace{2cm}
\begin{center}
{\bf {\large EXACT S-MATRICES FOR NONSIMPLY-LACED}} \\
\vspace{.1in}
{\bf {\large AFFINE TODA THEORIES }} \\
\vspace{1.5cm}
{G.W. DELIUS and
M.T. GRISARU}\footnote{On leave from Brandeis University, Waltham, MA
02254,
USA\\Work partially supported by the National Science Foundation under
grant PHY-88-18853 and by INFN.}\\
\vspace{1mm}
{\em Theory Division, CERN, 1211 Geneva 23, Switzerland}\\

\vspace{5mm}
and

\vspace{5mm}
{D. ZANON} \\
\vspace{1mm}{\em Dipartimento di Fisica dell' Universit\`{a} di Milano
and} \\
{\em INFN, Sezione di Milano, I-20133 Milano,
Italy}\\
\vspace{1.1cm}
{\bf{ABSTRACT}}
\end{center}
\bq
We derive exact, factorized, purely elastic scattering matrices for
the affine Toda theories based on the nonsimply-laced Lie algebras
\ta, \tb, \tc, \td, and \tg, as well as the superalgebras \tA{} and \tB.
\eq

\vfill

\begin{flushleft}
CERN-TH.6337/91\\
IFUM 413/FT\\
December 1991
\end{flushleft}
\end{titlepage}

\renewcommand{\thefootnote}{\arabic{footnote}}
\setcounter{footnote}{0}
\newpage

\sect{Introduction}

A few years ago, in the course of investigating perturbations of
conformal
field theories, Zamolodchikov considered a class of integrable
perturbations that lead to theories with massive excitations whose
natural description is in terms of their S-matrices \cite{1}. This
resulted in a
renewal of interest in S-matrices for integrable two-dimensional systems
and, because of connections of the original work
to the Lie algebra $E_8$ on one hand,
and to Toda theories on the other hand, a flurry of activity developed
around the construction of S-matrices for affine Toda theories based
on various Lie algebras. This construction was successfully
carried out for the case of affine Toda theories based on the
simply-laced
algebras $a_n^{(1)}$, $d_n^{(1)}$ and $e_{6,7,8}^{(1)}$, but,
except for $a_{2n}^{(2)}$,
failed
utterly for the families of nonsimply-laced algebras
$a_{2n-1}^{(2)}$, $b_n^{(1)}$,
$c_n^{(1)}$, $d_{n+1}^{(2)}$ as well as for $d_4^{(3)}$, $e_6^{(2)}$,
 $f_4^{(1)}$
and $g_2^{(1)}$ \cite{1.5,2,3}. Furthermore, extending the construction
to the case
of affine Toda theories based on those Lie superalgebras
which have  massive excitations only, we were able to
construct S-matrices for the Lie superalgebras $A^{(2)}(0,2n-1)$ and
$C^{(2)}(n+1)$ (whose bosonic root subsystem is simply-laced), but not
for $B^{(1)}(0,n)$ and $A^{(4)}(0,2n)$ \cite{4}. In this paper we remedy
this
situation.

Affine Toda  theories based on Lie algebras are massive two-dimensional
bosonic
field theories
represented by lagrangians of
the form
\EQ\label{1.1}
{\cal L}= -\frac{1}{2\b^2} \vec{\phi} \Box \vec{\phi} -\frac{\m^2
}{\b^2}
\sum_i q_i
 e^{\vec{\a}_i\cdot
 \vec{\phi}}
 \EN
where  the
$\vec{\a}_i$ are the simple roots of a rank $n$ Lie algebra augmented by
(the negative of) a maximal root and
$\vec{\phi}=(\phi_1,\phi_2,...\phi_n)$
are bosonic fields
describing $n$ massive particles. For the simply-laced theories all the
roots have the same length. (For the case of Lie superalgebras the set
of
roots is divided into bosonic and fermionic ones, and the lagrangians
contain in general fermions as well.) The Ka\v{c} labels $q_i$ are such
that
$\sum q_i \vec{\a}_i=0$ and the Ka\v{c} label corresponding to the
maximal
root is normalized to unity. We define $h=\sum_i q_i$ and refer to it,
somewhat imprecisely, as the Coxeter number. The
coupling constant is denoted by $\b$ and $\m$ sets the mass
scale. For notational convenience we will often omit these two
constants.

The Toda field equations may be viewed as integrability conditions
for a Lax pair. From this follows the existence of an infinite set of
currents $J_{\pm}^{(s)}$ of increasing spin $s$ (the first of which is
the
stress tensor) which
are conserved by virtue of the
field equations. Thus,
these theories are classically
integrable: the lagrangian above admits an infinite number of symmetries
described by the corresponding conserved charges.
At the quantum level, the existence of such symmetries has a profound
effect on the
structure of the scattering amplitudes of these theories: the
$n$-particle
S-matrices
factorize into a product
of elastic  two-particle S-matrices satisfying Yang-Baxter relations.
In principle, some additional assumptions of analyticity, unitarity and
bootstrap principle \cite{5} should allow them to be
determined exactly.

Elastic, unitary S-matrices for a process $a+b\rightarrow a+b$ can be
written as
products of ratios of hyperbolic
 sines
\beq\label{blocks}\label{1.2}
S_{ab}(\th)=\prod\bl{x}
\mbox{~~,~~}\qquad
\bl{x}\equiv{\sh\left({\th\over2}+{i\pi\over2h}x\right)
\over\sh\left({\th\over2}-{i\pi\over2h}x\right)}
\eeq
where $\th=\th_a-\th_b$ is the relative rapidity
and $h$ can be identified with the
Coxeter number of the Lie algebra. For $x<h ~mod~2h$ these S-matrices
have
physical sheet simple poles at $\th =\frac{i\pi x}{h}$ and
these can be interpreted as elementary particle poles from s-channel or
u-channel exchange, with masses related to the values of $x$.
These values are
constrained by the  bootstrap principle:
any physical sheet simple pole {\em must } be interpretable in
this manner and the corresponding particle must appear  in the
set $\{a,b,...\}$ which labels the S-matrix.
Values of $x$
outside the above range are not restricted. Coupling constant
dependence (with $S(\b =0) =1$) may be introduced by means of additional
blocks $(x \pm B)^{-1}$. Here $B(\b )$ (with $B(0) =0$) is {\em a
priori}
 arbitrary,
but there is good evidence that $B=
\frac{\b^2}{2\pi}(1+\frac{\b^2}{4\pi})^{-1}$.
Thus, typically
\EQ\label{1.3}
S_{ab}(\th ) = \prod_{x\in A_{ab}} \frac{(x)}{(x \pm B)}
\EN
for some set $A_{ab}$.
If the particles are self-conjugate, crossing symmetry
$S(\th ) = S( i\pi -\th )$ implies that along with the blocks
$(x)(x-B)^{-1}$
the
above product also contains the blocks $(h-x)(h-x+B)^{-1}$
The blocks in the numerator determine pole positions, while the blocks
in the denominator determine coupling constant dependence.

Additional restrictions come from the assumption of the S-matrix
bootstrap
which says that if $S_{ab}$ has a pole at $\th = \th_{ab}^c$
corresponding to the s-channel
exchange of a
particle $c$, then
\EQ\label{1.4}
S_{cd}(\th ) = S_{ad}(\th + \bar{\th}_{ac}^b)
S_{bd}(\th - \bar{\th}_{bc}^a)
\EN
where $\bar{\th}=i\pi-\th$.
The bootstrap principle and the S-matrix bootstrap
severely restrict the choice of s-channel  exchanges.
In the usual construction one requires \reff{1.4} to be satisfied
by the numerator of \reff{1.3} independently of the denominator and this
reduces the set $A_{ab}$ to integers, fixing
the mass ratios of the elementary particles to values independent of
the coupling constant.

For simply-laced Toda theories, the S-matrices are constructed
by first requiring tree-level consistency with the {\em classical} mass
spectra and with
the three-point couplings (as extracted from the Toda lagrangians)
which determine possible one-particle exchanges.
For example, for the $a_n^{(1)}$ affine Toda theory, with Coxeter number
$h=n+1$ and classical mass
spectrum
\EQ
m_a=2\m \sin \frac{a\pi}{h}    ~~~~~, ~~~~a=1,2,...,n
\EN
the proposed S-matrix is given by \cite{1.5}
\EQ
S_{ab}=\{a+b-1\}\{a+b-3\}...\{|a-b|+1\}
\EN
in terms of the convenient notation \cite{2}
\EQ
\{x\}=\frac{(x+1)(x-1)}{(x+1-B)(x-1+B)}
\EN
At the quantum level one checks that (one-loop) corrections
do not affect the mass spectrum
(aside from an overall rescaling that can be absorbed into the mass
scale). Other important checks consist in
correctly locating and identifying various higher-order poles as
anomalous threshold singularities \cite{2}.
Similar constructions exist for the other simply-laced cases and for
two theories based on Lie superalgebras.

However, such construction has failed
for affine Toda theories based on
nonsimply-laced algebras on several counts:
a) As mentioned above, the  bootstrap
principle leads to restrictions on the masses of the particles.
These restrictions are consistent with the classical masses, for both
simply-laced and nonsimply-laced theories, but for the nonsimply-laced
theories
radiative corrections distort the mass spectrum in a manner that at
first sight seems incompatible with bootstrap restrictions \cite{2,3,6}.
b) For the nonsimply-laced Toda theories an
explanation of higher-order poles as anomalous thresholds
has failed to account for all the singularities of proposed
S-matrices. c) As discussed in Refs. \cite{7,8} certain anomalous
threshold singularities lead to a breakdown of the charge
conservation rules used in, and implied by,  the bootstrap procedure.
These facts have led to speculations that factorizable, elastic
S-matrices
do not exist for these Toda theories, perhaps because the classical
integrability breaks down at the quantum level due to anomalies.

In Ref. \cite{8} we have demonstrated that at the quantum level, Toda
theories
for both simply-laced and nonsimply-laced algebras do have higher spin
conserved currents, albeit modified by quantum corrections. Therefore,
the difficulties encountered in attempts to
construct elastic, factorizable S-matrices for the
nonsimply-laced cases cannot be blamed on a failure of the charge
conservation laws.
What is required  is some modification of the usual construction
which takes into account the distortion of the mass spectrum,
the anomalous threshold structure, and, where it occurs, the breakdown
of the
bootstrap. As we shall demonstrate, one only has to give up the
idea that the blocks $(x)$ which determine particle pole positions
satisfy the bootstrap independently of the blocks $(x\pm B)$.
This allows $(x)$ to be coupling constant dependent and makes possible
the construction of satisfactory S-matrices.

Coupling constant dependence in the numerator blocks of \reff{1.3}
should not come as a surprise since quantum corrections do
change the mass spectrum. What is remarkable is the fact
that the procedure we will describe can be implemented consistently in
just a few
ways, and that the various
nonsimply-laced Toda theories take advantage of this. Thus
we are able to construct
exact factorized S-matrices for all the families of nonsimply-laced
Toda theories $a_{2n-1}^{(2)}$, $b_n^{(1)}$, $c_n^{(1)}$, $d_n^{(2)}$,
and
the \tg theory,
as well as the superalgebra cases $A^{(4)}(0,2n)$ and $B^{(1)}(0,n)$.
We leave the theories based on $d_4^{(3)}$, $e_6^{(2)}$ and
$f_4^{(1)}$ as an exercise for the reader.
As we will see, when the dust settles, these S-matrices
differ from previous ones in only one respect: the Coxeter number
which appears in the mass formulas and in the S-matrix blocks
has to be replaced by a ``renormalized'', coupling constant dependent
Coxeter number.

In the next sections we present the construction and  the
perturbative verifications. For general information about the
techniques used the reader should consult  some of the literature
in the references, in particular \cite{2,4,9}.
We will proceed in alphabetical order, from $a_{2n-1}^{(2)}$, to
$b_n^{(1)}$,
to $c_n^{(1)}$, etc.,
giving full details for the
first nonsimply-laced Toda theory and less for the
others, leaving some of the checks of the later theories
for the enterprising reader. A brief
account of the procedure was presented in Ref.~\cite{10}.

\sect{The S-matrix of the $a_{2n-1}^{(2)}$ theory}

The Toda theory
based on the nonsimply-laced twisted affine Lie algebras
$a_{2n-1}^{(2)}$
is described by the classical lagrangian
\bea\label{action}\label{2.1}
\b^2 {\cal{L}} &=& - {\frac12}\sum_1^{n-1} \phi_a
\Box \phi_a -{\frac12} \phi_n \Box \phi_n\\
&-&2 \sum_{k=2}^{n-1}  \exp{\left(-\sqrt{\frac{2}{h}} \sum_{1}^{n-1}
m_a \cos{\frac{(2k-1)a \pi}{h}} ~\phi_a \right)}
- \exp{\left( -\sqrt{\frac{2}{h}} \sum_{1}^{n-1} (-1)^a m_a \phi_a
\right)}\nonumber\\
&-& \exp{\left( -\sqrt{\frac{2}{h}} \sum_{1}^{n-1} m_a \cos{\frac{a \pi}
{h}} ~\phi_a + \phi_n \right)}
-  \exp{\left( -\sqrt{\frac{2}{h}} \sum_{1}^{n-1} m_a \cos{\frac{a\pi}
{h}} ~\phi_a - \phi_n \right)}\nonumber
\ena
with masses and
three-point couplings given by
\beq\label{m}\label{2.2}
m_n^2 = 2, \qquad m_a^2 = 4 m_n^2 \sin^2{(\frac{a\pi }{h})}
{}~~~~~~~a=1,...,n-1 ,
\eeq
\bea\label{c}\label{2.3}
c_{abc} &=& -\frac{\b}{\sqrt{2h}}m_a m_b m_c =
-\frac{4\b}{\sqrt{h}}\D_{abc}
{}~~~~~~~~~~~~{\rm{if}} ~~a+b+c=h
\nonumber \\
c_{abc} &=& \frac{\b}{\sqrt{2h}}m_a m_b m_c
=\frac{4\b}{\sqrt{h}}\D_{abc}
{}~~~~~~~~~~~~~~{\rm{if}} ~~a\pm b
\pm c=0
\nonumber \\
c_{ann} &=&\b \sqrt{\frac{2}{h}} m_n^2 m_a \cos{\frac{a \pi}{h}}
=\frac{4\b}{\sqrt{h}}\D_{ann}
\ena
All other three-point couplings are zero. Here $\D_{abc}$ represents the
area of a triangle with sides $m_a$, $m_b$, $m_c$.
We have introduced the Coxeter number $h=2n-1$ and set the overall
mass-scale
 $\m =1$.

In perturbation theory the only divergences one encounters in these
bosonic systems come from tadpoles. A finite quantum theory can be
defined
by absorbing such divergences into a renormalization of the mass scale
and, for these nonsimply-laced cases, a shift in the vacuum expectation
values of the fields which leads to a renormalization of the Ka\v{c}
labels.
The bare Ka\v{c} labels and the bare mass-scale are chosen such that the
quantum
lagrangian is identical in form to the classical lagrangian in
\reff{2.1}, but
with normal-ordered exponentials. As we will mention later on, a
slightly
different prescription is required when fermions are present.

At the one-loop level radiative corrections lead to mass shifts that
have been computed in Ref.~\cite{6}. They give
\beq\label{dm}\label{2.4}
\d  {m_a^2\over m_n^2}= -\frac{a\b^2}{h^2}\sin \frac{2a\pi}{h}.
\eeq
(In contrast,  for a simply-laced theory and also, exceptionally, for
the $a_{2n}^{(2)}$ theory
the mass ratios do not receive corrections.)
Without loss of generality we will write the radiatively-corrected
masses,
with
reference to \reff{2.2}, as
\beq\label{renmasses}\label{2.5}
\widetilde{m}_a^2=4 \widetilde{m}_n^2\sin^2\left({\pi\over
h}\left(a+\e_a(\b)\right)\right)
\eeq
At this stage, the $\e_a(\b )$ are in principle computable to any
order of perturbation theory.
The S-matrix will be defined with respect to these  masses.

In Ref. \cite{8} we have demonstrated the existence of a
quantum-conserved
spin 3 current for the $n=2$ case,
\bea\label{rJ}\label{2.6}
J_+^{(3)} &=& (1+\frac{\a}{2})(\pa_+\phi_1)^2(\pa_+\phi_2)^2
-\frac{\a}{12}(1+3\a+\a^2)(\pa_+^2\phi_1)^2
-\frac{\a}{12} (\pa_+\phi_1)^4 -\frac{\a}{12} (\pa_+\phi_2)^4
\nonumber\\
&&+(1+\frac{23}{12}\a +\a^2 +\frac{1}{6}\a^3)
(\pa_+^2\phi_2)^2+(2+3\a +\a^2)
\pa_+\phi_1\pa_+\phi_2\pa_+^2\phi_2
\ena
(with a suitable expression for $J_-^{(3)}$). Here
$\a=\frac{\b^2}{2\pi}$.
We expect that a similar conserved current exists for all the
$a_{2n-1}^{(2)}$ Toda theories.
Together with the renormalized stress tensor this defines corresponding
spin 1 and spin 3 charges
\EQ
Q^{(1)}=\int dx^+T_{++} ~~~,~~~Q^{(3)}=\int dx^+ J^{(3)}_+
\EN
whose presence guarantees the existence of factorizable
elastic S-matrices. The bootstrap procedure should then determine them.
However, as described in the above reference, the presence of certain
anomalous threshold singularities implies a breakdown of the usual
bootstrap results when
the $\langle\phi_a\, \phi_b\, \phi_{h-a-b}\rangle$ vertex
is involved. We will see later on precisely where and how this occurs.

\subsection{The S-matrix}

We construct the S-matrix by following procedures similar to those
used in the simply-laced case, but with two important differences: we
admit that the S-matrix has simple particle poles at positions shifted
away
from the classical mass values by radiative corrections, and we
relax the bootstrap
principle since some simple poles are shifted
away from their single-particle positions due to anomalous threshold
effects. (We will prove that all the shifted poles we
find can be accounted for in this fashion.)

Assuming that higher-spin quantum-conserved currents similar to those in
\reff{rJ} exist for all the $a_{2n-1}^{(2)}$ theories, we postulate
the existence of purely elastic two-body amplitudes.
We start by determining the element $S_{nn}$.
Since the three-point couplings in \reff{c} suggest that all the
particles
$\phi_1\cdots\phi_{n-1}$ appear as intermediate particles in this
process
the S-matrix bootstrap will allow the determination of all the other
amplitudes.
The intermediate particles lead to both
$s$- and $u$-channel poles
in $S_{nn}$.
In the rapidity plane these poles are located at:
\bea\label{Snnpoles}\label{2.8}
s-\widetilde{m}^2_a= 2\widetilde{m}_n^2(1+\ch\th)-\widetilde{m}_a^2=0
&&\Rightarrow\qquad\th={i\pi\over h}(h-2a-2\e_a)\nonumber\\
u-\widetilde{m}^2_a= 2\widetilde{m}_n^2(1-\ch\th)-\widetilde{m}_a^2=0
&&\Rightarrow\qquad\th={i\pi\over h}(2a+2\e_a)
\eea
We must reproduce these poles with the building blocks in
\reff{blocks}. The choice
\beq\label{Snnmin}
S_{nn}^{(min)}=\prod_{a=1}^{n-1}\ \bl{2a+2\e_a}\,\bl{h-2a-2\e_a}
\eeq
has the right pole structure. In order for the S-matrix to
reduce to the
identity matrix when the coupling constant $\b$ is zero we make the
crossing symmetric Ansatz
\beq\label{Snn}
S_{nn}=\prod_{a=0}^{n-1}\ {\bl{2a+2\e_a}\,\bl{h-2a-2\e_a}
\over\bl{2a+2\eta_a}\,\bl{h-2a-2\eta_a}}
\eeq
where $\e_a$ and $\eta_a$ are both zero when $\b=0$.
We let
the product start at $a=0$. This is
required for agreement with the tree level amplitudes.
None of the extra blocks which we
introduced in \reff{Snn} should produce  additional poles on the
physical
sheet (i.e., for $0<\th<i\pi$) and this requires
\beq\label{nopoles}
{h\over2}-a\geq\eta_a\geq
-a\qquad,\qquad \e_0\leq 0.
\eeq

The bootstrap principle  allows the
determination of the S-matrix elements $S_{an}$, \hbox{$a=1\cdots
n-1$}, through
\beq\label{bootSan}
S_{an}(\th
)=S_{nn}(\th+{1\over2}\th_{nn}^a)S_{nn}(\th-{1\over2}\th_{nn}^a)
\eeq
Here $\th_{nn}^a$ is the relative rapidity at which $S_{nn}$ has the
s-channel
pole corresponding to particle $a$, i.e., $\th_{nn}^a={i\pi\over
h}(h-2a-2\e_a)$, cf. \reff{Snnpoles}.
Using
\beq
\bl{x}_{(\th+{i\pi\over h}\r)}\cdot\bl{x}_{(\th-{i\pi\over h}\r)}=
\bl{x+\r}_{(\th)}\cdot\bl{x-\r}_{(\th)}
\eeq
we find
\bea\label{Sanlong}
S_{an}=\ \prod_{p=0}^{n-1}
&&{\bl{2p+2\e_p-{h\over2}+a+\e_a}\bl{h-2p-2\e_p-{h\over2}+a+\e_a}
\over\bl{2p+2\eta_p-{h\over2}+a+\e_a}\bl{h-2p-2\eta_p-{h\over2}+a+\e_a}}
\nonumber\\
\times&&{\bl{2p+2\e_p+{h\over2}-a-\e_a}\bl{h-2p-2\e_p+{h\over2}-a-\e_a}
\over\bl{2p+2\eta_p+{h\over2}-a-\e_a}\bl{h-2p-2\eta_p+{h\over2}-a-\e_a}}
\eea

This expression has a large number of poles, many more than perturbation
theory
can be expected to explain. However for special values of $\e_a$ and
$\eta_a$
many of the  blocks cancel each other, as we now show.
We rewrite the expression in \reff{Sanlong} by using the relations
$(x \pm 2h) =(x)$ and  $(-x) = (x)^{-1}$
and bring some of the blocks involving $\eta$ to the numerator.
We write
\EQ
S_{an} (\th ) = \frac{ N(\e , \eta )}{D( \e , \eta )}
\EN
where
\bea
N(\e , \eta ) &=& \prod_{p=0}^{n-1} (\frac{h}{2}
+2p-a+2\e_p-\e_a)(\frac{h}{2}
-2p+a-2\e_p+\e_a) \nonumber\\
 &\cdot&(\frac{h}{2} -2p-a-2\eta_p-\e_a)(\frac{h}{2} +2p+a+2\eta_p+\e_a)
\nonumber\\
&=& \prod_{p=n-a}^{n-1}(\frac{h}{2} -a -2p-2\eta_p-\e_a) \prod_{p=n-a}
^{n-1}(\frac{h}{2}+a+2p+2\eta_p+\e_a) \nonumber\\
&& \prod_{p=0}^{n-a-1}(\frac{h}{2} -a -2p-2\eta_p-\e_a) \prod_{p=a}
^{n-1}(\frac{h}{2}+a-2p-2\e_p+\e_a) \nonumber\\
&& \prod_{p=1}^{a}(\frac{h}{2} -a +2p+2\e_p-\e_a) \prod_{p=0}
^{a-1}(\frac{h}{2}+a-2p-2\e_p+\e_a) \nonumber\\
&& \prod_{p=a+1}^{n-1}(\frac{h}{2} -a +2p+2\e_p-\e_a) \prod_{p=1}
^{n-a-1}(\frac{h}{2}+a+2p+2\eta_p+\e_a) \nonumber\\
&&\cdot (\frac{h}{2}-a+2\e_0-\e_a)(\frac{h}{2}+a+2\eta_0+\e_a)
\ena
and $D(\e , \eta )$ is obtained by the interchange $2\e_p
\leftrightarrow 2\eta_p$ in the above expression.
After making suitable changes of the variable $p$ in each of the
above products we can rewrite
\bea\label{2.17}
N(\e , \eta ) &=& \prod_{p=n-a}^{n-1} (\frac{h}{2} -a -2p -2\eta_p
-\e_a)
(\frac{h}{2} -a -2p +2\eta_{h-a-p}+\e_a) \nonumber\\
&&\prod_{p=1}^{a-1} (\frac{h}{2} -a -2p -2\eta_p -\e_a)
(\frac{h}{2} -a -2p -2\e_{a+p}+\e_a) \nonumber\\
&&\prod_{p=1}^{a-1} (\frac{h}{2} -a +2p +2\e_p -\e_a)
(\frac{h}{2} -a +2p -2\e_{a-p}+\e_a) \nonumber\\
&&\prod_{p=a+1}^{n-1} (\frac{h}{2} -a +2p +2\e_p -\e_a)
(\frac{h}{2} -a +2p +2\eta_{p-a}+\e_a) \nonumber\\
&&\cdot (\frac{h}{2}-a+2\e_0-\e_a)(\frac{h}{2}+a+2\eta_0+\e_a)
(\frac{h}{2}-a-2\eta_0-\e_a)\nonumber\\
&&\cdot (\frac{h}{2}-a-\e_a) (\frac{h}{2}+a+\e_a)
(\frac{h}{2}+a-2\e_0+\e_a)
\ena
{\samepage with a corresponding expression for $D(\e , \eta )$:
\bea\label{2.18}
D(\e , \eta ) &=& \prod_{p=n-a}^{n-1} (\frac{h}{2} -a -2p -2\e_p -\e_a)
(\frac{h}{2} -a -2p +2\e_{h-a-p}+\e_a) \nonumber\\
&&\prod_{p=1}^{a-1} (\frac{h}{2} -a -2p -2\e_p -\e_a)
(\frac{h}{2} -a -2p -2\eta_{a+p}+\e_a) \nonumber\\
&&\prod_{p=1}^{a-1} (\frac{h}{2} -a +2p +2\eta_p -\e_a)
(\frac{h}{2} -a +2p -2\eta_{a-p}+\e_a) \nonumber\\
&&\prod_{p=a+1}^{n-1} (\frac{h}{2} -a +2p +2\eta_p -\e_a)
(\frac{h}{2} -a +2p +2\e_{p-a}+\e_a) \nonumber\\
&&\cdot (\frac{h}{2}-a+2\eta_0-\e_a)(\frac{h}{2}+a+2\e_0+\e_a)
(\frac{h}{2}-a-2\e_0-\e_a)\nonumber\\
&&\cdot (\frac{h}{2}-a-2\eta_a+\e_a) (\frac{h}{2}+a+2\eta_a-\e_a)
(\frac{h}{2}+a-2\eta_0+\e_a)
\eea}

We observe that one choice which cancels many poles is
$\e_a=0$ and $\eta_a={B\over2}$ for all $a$ and some $B=B(\b)$.
Then we are left
with
\beq
S_{an}^{\rm (unren)}=\prod_{p=1}^{2a-1}\ {\bl{{h\over2}-a+p}^2\over
\bl{{h\over2}-a+p+B}\,\bl{{h\over2}-a+p-B}}\
{\bl{{h\over2}-a}\over\bl{{h\over2}-a+B}}\,
{\bl{{h\over2}+a}\over\bl{{h\over2}+a-B}}
\eeq
This choice corresponds to unrenormalized mass ratios and is not
relevant
here.

However there is a choice which has even fewer poles: we require
\bea
-2\eta_p-\e_a=  2\e_{h-a-p}+\e_a &\rightarrow& \e_{h-a-p}=-\e_a-\eta_p
\nonumber\\
2\eta_{h-a-p}+\e_a = -2\e_p-\e_a &\rightarrow& \eta_{h-a-p}=-\e_a-\e_p
\ena
for $p=n-a,...,n-1$, which entirely cancels the first line in
$N(\e,\eta)$
against that in $D(\e,\eta)$. Similarly, requiring
\bea
-2\eta_p-\e_a=-2\eta_{a+p} +\e_a &\rightarrow& \eta_{a+p}=\e_a+\eta_p
\nonumber\\
-2\e_{a+p}+\e_a =-2\e_p-\e_a &\rightarrow& \e_{a+p}=\e_a+\e_p
\nonumber\\
2\e_p-\e_a=2\e_{p-a}+\e_a &\rightarrow& \e_{p-a}=\e_p-\e_a \nonumber\\
2\eta_p-\e_a=2\eta_{p-a}+\e_a &\rightarrow&\eta_{p-a}=\eta_p-\e_a
\ena
cancels the second and fourth lines. The above equations have the
general
solution
\EQ
\e_a=a\e  ~~~~,~~~~\eta_a=(a-h)\e ~~~~~~~~~a=1,2,\cdots ,n-1
\EN
for some arbitrary $\e (\b )$ with $\e(0)=0$.
Additional poles from the last two lines in \reff{2.17} may be cancelled
by
choosing
\EQ
\e_0=0
\EN
and
\EQ\label{2.24}
\eta_0 = -h\e      ~~~~{\rm or} ~~\eta_0=0
\EN
However, the second choice for $\eta_0$ would completely cancel the
$a=0$ block
in \reff{Snn} and we would fail to account at the tree level for the
part of
$S_{nn}$ which is independent of $\th$ (cf. $B(\th)$ in eq.(2.5)
of Ref.~\cite{4}. The theory considered there has the same bosonic tree
level
amplitude as the present theory). The first choice for $\eta_0$ gives
the
description of the S-matrix for the theory at hand. As we discuss later,
the second choice leads eventually to the S-matrix
for the Toda theory based on the Lie superalgebra $A^{(4)}(0,2n)$, where
a part independent of $\th$ is indeed absent.
We note that $\e(\b)$ satisfies $-\frac{1}{2n}\leq\e\leq 0$ in order to
fulfill \reff{nopoles}.

The S-matrix element $S_{an}$ becomes
\bea\label{San}
S_{an}=\prod_{p=1}^{a-1}&&{\bl{{h\over2}+(-a+2p)(1+\e)}^2\over
\bl{{h\over2}+(-a+2p)(1+\e)-2h\e}\,\bl{{h\over2}+(-a+2p)(1+\e)+2h\e}}
\nonumber\\
\times&&
{\bl{{h\over2}-a(1+\e)}\over\bl{{h\over2}-a(1+\e)-2h\e}}\,
{\bl{{h\over2}+a(1+\e)}\over\bl{{h\over2}+a(1+\e)+2h\e}}
\eea
This
expression still has two simple poles at $\th={i\pi\over
h}\left({h\over2}+a(1+\e)\right)$ and at $\th={i\pi\over
h}\left({h\over2}-a(1+\e)\right)$ as well as several double poles. We
discuss
the double poles later on and note here that the simple poles
correspond to particle $\phi_n$ in the intermediate s- and u-channels.
Indeed
\beq
s-\widetilde{m}_n^2=\widetilde{m}_n^2+\widetilde{m}_a^2+2\widetilde{m}_n
\widetilde{m}_a\ch\th
-\widetilde{m}_n^2=0\qquad\mbox{at }\ \th={i\pi\over
 h}\left({h\over2}+a(1+\e)\right)
\eeq
and similarly for the u-channel.

It is convenient to introduce some simplifying notation:
\EQ
B=-2h{\e\over 1+\e} ~~~~,~~~H=\frac{h}{1+\e } =h+\frac{B}{2}
\EN
\EQ
\blt{x}={\sh\left({\th\over2}+{i\pi\over2\newh}x\right)
\over\sh\left({\th\over2}-{i\pi\over2\newh}x\right)}\qquad ,\qquad
\cbl{x}={\blt{x-1}\blt{x+1}\over\blt{x-1+B}\blt{x+1-B}}
\EN
Therefore \reff{Snn} and \reff{San} become
\bea\label{nSnn}
S_{nn}&=&\prod_{a=0}^{n-1}\ {\blt{2a}\,\blt{\newh-2a}
\over\blt{2a+B}\,\blt{\newh-2a-B}}\\
\label{nSan}
S_{an}&=&\prod_{p=1}^a\ \cbl{{\newh\over2}+2p-a-1}=S_{na}.
\eea
In terms of this notation \reff{renmasses}  implies that the
renormalized
masses have the same form as the classical masses, but
with $h$ replaced by $\newh$.

The remaining S-matrix elements $S_{ab},\ a,b=1\cdots n-1$ are obtained
by
another application of the bootstrap
\bea\label{bootSab}
S_{ab}(\th)&=&S_{nb}\left(\th+\shalf\th_{nn}^a\right)
S_{nb}\left(\th-\shalf\th_{nn}^a\right)\nonumber\\
&&=\prod_{p=1}^b\ \cbl{2p-b-1+a}\,\cbl{\newh+2p-b-1-a}\label{nSab}
\eea
This expression is symmetric in $a,b$ as one can verify by using
relations
such as $\prod_{p=0}^x\cbl{2p-x}=1$. Crossing symmetry can be
easily checked by a change of variable $p\rightarrow-p+b+1$ in one of
the terms.

$S_{ab}$ has four simple poles. For $a>b$, these are located at
${\newh\over
i\pi}\th=(a-b),(\newh-a+b),(\newh-a-b)$, $(a+b)$. From the couplings in
the lagrangian one expects that they could be identified with s-channel
and u-channel exchanges of the particles $\phi_{(a-b)}$ and
$\phi_{(a+b)}$ for $a+b < \frac{h}{2}$ or $\phi_{(h-a-b)}$ for
$a+b>\frac
{h}{2}$.
If the pole at
$\th={i\pi\over\newh}(a-b)$ corresponds to a particle in the u-channel
then this
particle has mass
\beq
\mt^2=\mt_a^2+\mt_b^2-2\mt_a\mt_b\cos{\pi\over\newh}(a-b)=
4\mt_n^2\sin^2{(a-b)\pi\over\newh}
\eeq
and this identifies it as particle $\phi_{(a-b)}$. Similarly the pole at
$\th={i\pi\over\newh}(\newh -a+b)$ is identified as the s-channel pole
of
the same
particle.

The same calculation for the pole at $\th={i\pi\over\newh}(a+b)$ leads
to a mass
\beq\label{massapb}\label{2.33}
\mt^2=\mt_a^2+\mt_b^2+2\mt_a\mt_b\cos{\pi\over\newh}(a+b)=
4\mt_n^2\sin^2{(a+b)\pi\over\newh}
\eeq
If $a+b<{h\over2}$ this is the mass of particle $\phi_{(a+b)}$. But if
$a+b>{h\over2}$ the pole does not appear at the expected position
which should correspond, according to the couplings in \reff{c},
to the particle $\phi_{(h-a-b)}$. This particle
has a mass
\beq
\mt^2_{(h-a-b)}=4\mt_n^2\sin^2{(h-a-b)\pi\over\newh}
\eeq
which is not equal to that in \reff{massapb} due to the fact that $h\neq
\newh$.
As we will explain in subsection \ref{shifts}, this displacement of the
pole and
the attendant breakdown of the bootstrap principle is
due to the presence of an
anomalous threshold singularity.

Aside from some information about the particle content and the
existence of particular three-point couplings, as well as the fact that
the masses renormalize in a nontrivial manner, we have used very
little information about the specific Toda lagrangian. Some
choices were made in the course of the construction of the S-matrix:
no extra CDD poles and zeroes in $S_{nn}$, and a minimal number of
poles in $S_{an}$. After that, the solution of the bootstrap was
essentially unique. However, it is necessary to check the
self-consistency
of the result, and to show that the S-matrix indeed describes
scattering in the particular Toda theory.
Consistency amounts to showing that the bootstrap closes on the
set of particles we have introduced (or explaining where and why it
fails), and
in identifying all the higher-order poles as anomalous threshold
singularities.
Contact with the Toda lagrangian consists in checking tree-level
agreement, and
agreement at the loop-level for masses, coupling constant dependence,
and coefficients of the various poles. In particular we must determine
the relation between the only free parameter which the S-matrix
contains,
namely $B$, and the coupling constant $\b$ of the Toda lagrangian.
We expect, from the experience with the simply-laced cases, that
$B= \frac{\b^2}{2\pi} \left( 1+\frac{\b^2}{4\pi} \right)^{-1}$ and we
will verify that this is indeed consistent with our results.

In carrying out the checks, we have a choice between using the classical
masses or the physical  masses. In fact it is more convenient to make
the
second choice, since our S-matrix is already expressed in terms of (we
believe) the  correct  masses. In this spirit, we would
rewrite the lagrangian in terms of these masses and counterterms,
and choose the counterterms to appropriately cancel part of the loop
corrections to the self-energy.
In principle we could do the same for the coupling constants, but to the
order to which we are working this will not be necessary.

To conclude this subsection
we note that our S-matrix, as well as the predicted
renormalized mass spectrum, have the conventional forms encountered in
the simply-laced theories, with the only difference that the Coxeter
number $h$ must be replaced everywhere with the {\em renormalized}
Coxeter number $H$.

\subsection{Consistency of the bootstrap}

We have used the single particle poles in $S_{nn}$ to determine all the
other
matrix elements by bootstrap. We have also identified some single
particle
poles in these matrix elements and these too could be used in the
bootstrap
leading to
the following consistency conditions:
\bea\label{bootstrapeqs}
S_{nn}(\th)&=&S_{an}(\th+\bar{\th}_{an}^n)
S_{nn}(\th-\bar{\th}_{nn}^a)\nonumber\\
S_{nb}(\th)&=&S_{ab}(\th+\bar{\th}_{an}^n)
S_{nb}(\th-\bar{\th}_{nn}^a)\nonumber\\
S_{(a-b)n}(\th)&=&S_{an}(\th+\bar{\th}_{a(a-b)}^b)
S_{bn}(\th-\bar{\th}_{b(a-b)}^a)\\
S_{(a-b)c}(\th)&=&S_{ac}(\th+\bar{\th}_{a(a-b)}^b)
S_{bc}(\th-\bar{\th}_{b(a-b)}^a)\nonumber\\
S_{(a+b)n}(\th)&=&S_{an}(\th+\bar{\th}_{a(a+b)}^b)
S_{bn}(\th-\bar{\th}_{b(a+b)}^a)\qquad\mbox{~~~}a+b<{h\over2}\nonumber\\
S_{(a+b)c}(\th)&=&S_{ac}(\th+\bar{\th}_{a(a+b)}^b)
S_{bc}(\th-\bar{\th}_{b(a+b)}^a)\qquad\mbox{~~~ }a+b<{h\over2}\nonumber
\eea
where $\bar{\th}_{ab}^c=i\pi-\th_{ab}^c$ and $\th_{ab}^c$ is the
location of
the single particle s-channel pole due to particle $c$ in $S_{ab}$:
\beq
\bar{\th}_{an}^n={i\pi\over\newh}({\newh\over2}-a),\ \:\:
\bar{\th}_{nn}^a={i\pi\over\newh}2a,\ \:\:
\bar{\th}_{a(a-b)}^b=\bar{\th}_{a(a+b)}^b={i\pi\over\newh}b.
\eeq
In writing the bootstrap equations \reff{bootstrapeqs} we have used the
fact
that
\beq
\th_{ab}^c+\th_{ca}^b+\th_{bc}^a=2\pi i\qquad\mbox{for }a,b,c=1\cdots n.
\eeq

The first equality in \reff{bootstrapeqs} is a simple consequence of
\reff{bootSan} and the fact that $S(\th+\pi i)S(\th)=1$. Similarly the
second
equation is a consequence of \reff{bootSab}, the 3rd of the 5th and the
4th of
the 6th. The last two equations are real consistency conditions on the
bootstrap.
They can be checked to be satisfied.

Notice the restriction to
$a+b<{h\over2}$. No bootstrap equations arise from the poles in $S_{ab}$
at
$\th={i\pi\over\newh}(a+b)$ and $\th={i\pi\over\newh}(h-a-b)$ if
$a+b>{h\over2}$
because these cannot be identified as single particle poles.

\subsection{Perturbative checks and coupling constant dependence}

We observe that the $a_{2n-1}^{(2)}$ lagrangian has the same form,
classical
mass spectrum and couplings as the bosonic part of the Toda lagrangian
for the
Lie superalgebra $A^{(2)}(0,2n-1)$ and therefore the bosonic
tree-level S-matrices
are the same. Furthermore, in both theories the exact expressions for
$S_{an}(\th )$ and $S_{ab}(\th )$  are identical, except for the
replacement of $h$ by $H$. Therefore, since $H=h+O(\b^2)$, and
having checked in our previous work \cite{4} that one obtains the
correct
result at the tree-level, no further checks are needed for these
S-matrix elements. On the other hand $S_{nn}$ is different in the
two theories, and a separate check is necessary to show that the
expression in \reff{nSnn} has the correct tree level value. This is
straightforward, and indeed we find agreement provided
$B=\frac{\b^2}{2\pi}+O(\b^4)$.
In this respect, extending the product in \reff{Snn} to
include the $a=0$ term was crucial.

One-loop checks allow us
to determine the coupling constant dependence to order $\b^4$.
This is achieved by computing at the one-loop level from the Toda
lagrangian
the residue of a suitable S-matrix element at a single particle pole and
comparing it to the corresponding residue extracted from the exact
S-matrix.
We find agreement provided $B={\b^2\over 2\pi}-{\b^4\over 8 \pi^2}$,
i.e. to
this order
\EQ
B=\frac{1}{2\pi}\frac{\b^2}{1+\frac{\b^2}{4\pi}}
\EN
which is the standard
dependence on the coupling constant that one has found in simply-laced
theories
\cite{11}. We give details of the calculation in Appendix B.

We observe that our S-matrix predicts a very specific
manner in which radiative effects shift the masses of the theory:
\EQ
\widetilde{m}_a^2 = 4 \widetilde{m}_n^2 sin^2\left( \frac{\pi a}{H}
\right) \EN
Using $H=h+{B\over2}=h+\frac{\b^2}{4\pi} +O(\b^4)$
this is consistent with the one-loop mass corrections in \reff{dm}, and
already provides a nontrivial check of our S-matrix and of the
restrictions that follow from the bootstrap. It will be
interesting to perform a direct two-loop calculation of the mass shifts
and compare with those predicted here.

As a further perturbative check, we have computed for the $a_3^{(2)}$
 case,
up to
one-loop order, the spin 3 charges of the two particles of the theory
\cite{7,8}, and found agreement with those predicted by the exact
S-matrix, see
\reff{8.6}.
Additional checks are obtained when we calculate the coefficients of
the anomalous threshold singularities in the next subsection.

\subsection{Anomalous threshold singularities and the shifted poles}
\label{shifts}
As we have mentioned, in comparing the exact S-matrix to perturbation
theory we have a choice between using radiatively corrected physical
masses (and
counterterms which would just remove the corresponding contributions
from self-energy diagrams), or classical  masses. In a strict
perturbation theory context the latter choice might be more logical,
and we would expand everything in powers of the coupling constant.
However, it is more economical to make the former choice, since it
avoids having to expand denominators such as $(s-\widetilde{m}^2)^{-1}$.
Furthermore, in a pure S-matrix context questions about the consistency
of the bootstrap and identification of higher order poles should be
possible in terms of the  masses of the asymptotic
fields. In any event, choosing to work with physical, quantum-corrected
masses
reveals further miraculous ways in which these Toda theories solve
their own problems.

In perturbation theory,  using physical  masses, (or in
a pure analyticity-unitarity S-matrix approach)
the amplitude $S_{ab}$
for $a+b >n$ would have not only a simple pole corresponding to the
particle
$\phi_{h-a-b}$  as shown in Fig. 1.a, but also neighboring poles
produced as
anomalous threshold singularities from the various diagrams in Fig.
1.b,c,d,e,f.
\begin{figure}
\vspace{130mm}\hspace{10mm}
\special{picture fig1}
\caption{Contributions to the $S_{ab}$ amplitude of the $a_{2n-1}^{(2)}$
theory which are responsible for the shift in the particle pole
position.}
\end{figure}
Indeed,
using the value of the physical  masses it is straightforward to check
by means
of a dual diagram analysis that the triangle diagrams produce  pole
singularities
located at $\th =\frac{i\pi}{\newh}(a+b)$
and the crossed box in Fig. 1.d has a double pole at the same position.
Specifically, denoting by $\d$ the shift of the S-matrix pole from its
expected position, i.e.
\EQ\label{2.41}
\d = 4\widetilde{m}^2_n \left( \sin^2
\frac{\pi}{\newh}(a+b) -\sin^2 \frac{\pi}{\newh}(h-a-b) \right)
\EN
and
$
\s =s-\widetilde{m}^2_{h-a-b}
$
we have the contributions from the six diagrams
\bea
(a) &:& ~~~\frac{1}{\s} c^2_{ab(h-a-b)} \nonumber\\
(b) &:& ~~~\frac{T_{ab}}{\s
-\d}c_{ann}c_{bnn}c_{abnn} \nonumber\\
(c) &:&~~~\frac{T_{ab}}{\s (\s -\d
)}c_{ann}c_{bnn}c_{ab(h-a-b)}c_{nn(h-a-b)}
\nonumber\\
(d) &:& ~~~\frac{R_{ab}}{(\s - \d )^2} c^2_{ann}c^2_{bnn} \nonumber\\
(e) &:& ~~~\frac{T^2_{ab}}{(\s -\d )^2} c^2_{ann}c^2_{bnn}c_{nnnn}
\nonumber\\
(f) &:& ~~~\frac{T^2_{ab}}{\s (\s -\d )^2} c^2_{ann}c^2_{bnn}
c^2_{nn(h-a-b)}
\eea
Here $T_{ab}$ and $R_{ab}$ are the simple
and double pole coefficients, and the coupling constants,
in a lowest order computation, can be obtained
from the lagrangian. For a complete comparison higher-order
corrections should be included, as well as contributions from the
subleading
parts of the triangle and crossed box diagrams.

The exact S-matrix has a simple pole at $\s = \d$.
As we describe in Appendix A
it is straightforward to check,  using standard values of $T_{ab}$
and $R_{ab}$ and the above value for
$\d$ as well as the values of the coupling constants, that in the
sum of the contributions (a), $2\times$(c), (d) and (f) the pole at
$\s=0$
 cancels.
Furthermore, to $ O(\b^4)$
the double pole  cancels as well,
leaving a  simple pole at $\s =\d$ with the correct
$O(\b^2)$ residue
and indeed reproducing the result obtained from the exact S-matrix.
The diagrams in Fig. 1.b,e contribute only
at higher order.
Thus, due to a subtle interplay between the location of the anomalous
threshold poles and their residues, together with the value of the
mass corrections,
the simple particle pole in the S-matrix
located
at the (renormalized) particle  mass has been cancelled, and replaced by
simple pole at the location of the
anomalous threshold singularity. Clearly, ordinary bootstrap ideas
should
not be applied in such a case and indeed, as we discussed in Refs.
\cite{7,8},
the charge conservation conditions for the corresponding vertex
function $\langle\phi_a \phi_b \phi_{h-a-b}\rangle$ which would follow
from the
ordinary bootstrap do not hold.

It is interesting to contrast the above discussion of the anomalous
poles with
that in the $A^{(2)}(0,2n-1)$ theory \cite{4} which differs from our
theory by
the addition of one fermion: there, the bosonic field $\phi_n$ gives
rise to the
same anomalous threshold structure, but this is precisely cancelled by
an
identical structure from the fermion, so that the simple particle pole
ends up
in the position predicted by the mass formula (of the classical theory)
and the
usual bootstrap applies.

We turn now to the  other anomalous threshold singularities
and their identification with higher order poles in the S-matrix.
As an  aid to further discussion we note that many of the results and
expressions derived in Ref.~\cite{4} for  some of the amplitudes
and the dual diagram constructions in
the $A^{(2)}(0,2n-1)$ theory can be taken over to our theory. Also,
as
mentioned above, the exact S-matrix
elements  $S_{an}$ and $S_{ab}$ have  the same form  in the two
theories,
provided the blocks are reinterpreted  according to our definitions
above
i.e. with $h \rightarrow \newh$.

We consider first the double poles in the amplitude $S_{an}$ which occur
at
\EQ
\th =\frac{i\pi}{\newh}\left(\frac{\newh}{2} +2p-a\right)~~~~~~~
p=1,2,...a-1
\EN
In Ref.~\cite{4} we encountered similar double poles in the
corresponding
$S_{an}$ amplitudes, and they were
accounted for by
both uncrossed and crossed box diagrams with dual diagrams such as those
of
Fig. 12 of that reference. Similar dual diagrams can be constructed in
the present theory. Remarkably, although the construction is done
with the renormalized masses, so that
the lengths of the lines are  different from what they were in the above
reference, nonetheless the dual diagrams exist for both the crossed and
the uncrossed boxes. The only change is in the location of the double
poles, and this agrees with their location in the exact S-matrix.
Indeed, let us consider the dual diagram in Fig. 2.
\begin{figure}
\vspace{65mm}\hspace{35mm}
\special{picture fig2 scaled 700}
\caption{Dual diagram for the double poles in $S_{an}$ at
$\th={i\pi\over H}\left({H\over2}+2p-a\right)$.}
\end{figure}
We observe that in a triangle with sides $p,n,n$
(i.e. with lengths given by $\widetilde{m_p}, \widetilde{m_n},
\widetilde{m_n}$,)the angle opposite the side $p$ is $\th_p =\frac{2\pi
p}{H}$.
This, and some
elementary Euclidean geometry is sufficient to show that the dual
diagram is planar, and that it predicts a double pole at
the desired location. Additional contributions at the same location
come from a crossed-box
diagram.
To one-loop order the
coefficients of these double poles are the same as for the superalgebra
case (where these contributions are also from bosons only),
and  since our $S_{an}$ has the same structure there is no
need for any further checks at the one-loop level.

A new feature appears when we examine the double poles of the $S_{ab}$
amplitudes. For example, in the superalgebra case one obtained double
poles from both an uncrossed and a
crossed box, cf. Fig. 13 of reference \cite{4} for the case of the
$S_{n-1,n-1}$
amplitude. Now however, using the
renormalized masses, the dual diagram for the uncrossed box can no
longer be drawn in the plane and it would seem that although a double
pole is still produced by the crossed box its coefficient would not
match that of the exact S-matrix; the contribution from the uncrossed
box
is needed. The explanation is provided by realizing that in addition to
the box diagram there are seemingly higher-order diagrams where one of
the
internal lines is replaced by the set of diagrams appearing in Fig. 1
which are responsible for the displacement of the single particle pole.
In going
through the Landau-Cutkoski analysis for the location and nature of
singularities, one realizes then that a double pole is indeed produced,
with the
correct residue. Equivalently one can perform  the dual diagram analysis
using
not the actual particle mass $\widetilde{m}^2_{h-a-b}$ but that
corresponding to the  pole of the S-matrix,
namely $\widetilde{m}^2_{h-a-b} + \d$. In this manner all the double
poles
of the exact S-matrix can be accounted for, and aside from verifying
that the
higher-order coefficients are correctly given we claim to have checked
the
self-consistency of the S-matrix we have constructed.

\sect{The $A^{(4)}(0,2n)$ theory}

The classical lagrangian for the
Toda theory based on the Lie superalgebra $A^{(4)}(0,2n)$ is
obtained from the lagrangian of the $a_{2n-1}^{(2)}$ theory
by dropping the field $\phi_n$ and adding a fermion:
{\samepage \bea
\b^2 {\cal{L}} &=& - {\frac12}\sum_1^{n-1} \phi_a
\Box \phi_a +\frac{i}{2}\bar{\psi}\g \cdot \pa \psi \\
&-& 2\sum_{k=2}^{n-1}  \exp{\left(-\sqrt{\frac{2}{h}} \sum_{1}^{n-1}
m_a \cos{\frac{(2k-1)a \pi}{h}} ~\phi_a \right)}
- \exp{\left( -\sqrt{\frac{2}{h}} \sum_{1}^{n-1} (-1)^a m_a \phi_a
\right)}\nonumber\\
&-&2 \exp{\left( -\sqrt{\frac{2}{h}} \sum_{1}^{n-1} m_a \cos{\frac{a
\pi}
{h}} ~\phi_a  \right)}
- \frac{1}{\sqrt{2}} \bar{\psi}\psi
 \exp{\left( -\frac{1}{\sqrt{2h}} \sum_{1}^{n-1} (-1)^a m_a ~\phi_a
\right)}\nonumber \ena}
with bosonic masses and
three-point couplings as in eq.(\ref{2.2},\ref{2.3}), $m_{\psi}^2=2$ and
$h=2n-1$.

In the quantum theory normal-ordering is not sufficient to remove
all infinities since, e.g., one-loop fermionic diagrams also lead to
divergences. It appears that the correct prescription (which can be
understood
also by observing that such theories can be thought of as theories with
explicitly broken supersymmetry \cite{12}) is to normal-order the first
and third exponentials, normal-order the exponential multiplying the
fermions, and write the second exponential (which in the supersymmetric
theory comes
from eliminating an auxiliary field and is the partner of the last term)
as the
square of a normal-ordered exponential.

We recall that in the theory for the Lie superalgebra $A^{(2)}(0,2n-1)$
which is obtained from the bosonic theory by adding the fermion but
without dropping the $\phi_n$ boson, the one-loop mass corrections
vanish.
Here, having dropped the boson and introduced the fermion, the
mass ratios $(m_a/m_{\psi})^2$ receive one-loop
corrections which have the same magnitude as those of the
$a^{(2)}_{2n-1}$
theory in \reff{2.4}, but
opposite sign.

One starts the S-matrix bootstrap by postulating an expression for
the fermion-fermion amplitude. Based on the previous case, and
using the fact that the fermions couple to all the other particles in
a manner similar to the couplings of $\phi_n$ we
are led to
\EQ
S_{\psi \psi} = - \prod_{a=1}^{n-1} \frac{(2a)_H (H-2a)_H}
{(2a-B)_H(H-2a+B)_H}
\EN
with
\EQ
H=h-\frac{B}{2} ~~~~,~~~~B=\frac{\b^2}{2\pi}\frac{1}
{1+\frac{\b^2}{4\pi}}
\EN

Some comments are in order. The above expression is very similar to the
one for $S_{nn}$ in the $a_{2n-1}^{(2)}$ theory and in fact one goes
through
exactly the same bootstrap procedure as before, with the following
changes: we have put a minus sign in front of the product, to account
for the fermionic nature of the particles, and
we have dropped the block corresponding to $a=0$. In the bosonic theory
that block was necessary essentially to account for a contribution to
the S-matrix independent of $\th$ and led us to choosing $\eta_0 = -h\e$
in \reff{2.24}.
Here we make instead the choice $\eta_0=0$ so that the S-matrix
vanishes at $\th =0$ (cf. the expression for the
fermion amplitude $F(\th )$ in eq.(2.5) of Ref.~\cite{4}). This choice
leads to
$H=h-\frac{B}{2}$ rather than $H=h+\frac{B}{2}$ which also agrees with
the
perturbation theory result that the mass corrections have an opposite
sign.
Beyond $O(\b^2)$ the above expression for $B$ is conjectural.

The rest of the S-matrix elements are obtained again by standard
bootstrap.
We find that
$S_{a\psi}$ is identical in form to the $S_{an}$ of the \ta theory
and so is $S_{ab}$ (with the different value of $H$ however). As in the
bosonic theory, for $a+b>n$, $S_{ab}$ has a simple pole shifted from the
single-particle position to an anomalous threshold position. This time
the relevant anomalous threshold singularity comes from a fermion loop
(instead of the loop containing $\phi_n$) and the diagram in Fig. 1e is
absent.
The pole shift is
somewhat different because of the change in sign implied by the new
value of $H$. However, using the fact (which follows from cancellations
between $\phi_n$ and $\psi$ contributions in the $A^{(2)}(0,2n-1)$
theory)
that now $\psi$ contributes $-T_{ab}$ and $-R_{ab}$
in \reff{2.41} while to lowest order
the sign of $\d$ changes, it is obvious that to $O(\b^4)$ at least
matters work in the same manner as
for the bosonic theory. The various double poles can
again be accounted for as before.

\sect{The $b_n^{(1)}$ theory}

The Toda theory
based on the nonsimply-laced  affine Lie algebras $b_n^{(1)}$
is described by the lagrangian
{\samepage \bea
\b^2 {\cal{L}} &=& - {\frac12}\sum_1^{n-1} \phi_a
\Box \phi_a -{\frac12} \phi_n \Box \phi_n\\
&-&2 \sum_{k=2}^{n}  \exp{\left(-\sqrt{\frac{2}{h}} \sum_{1}^{n-1}
m_a \cos{\frac{(2k-1)a \pi}{h}} ~\phi_a \right)} \nonumber\\
&-& \exp{\left( -\sqrt{\frac{2}{h}} \sum_{1}^{n-1} m_a \cos{\frac{a \pi}
{h}} ~\phi_a + \phi_n \right)}
-  \exp{\left( -\sqrt{\frac{2}{h}} \sum_{1}^{n-1} m_a \cos{\frac{a\pi}
{h}} ~\phi_a - \phi_n \right)}\nonumber
\ena}
The Coxeter number is $h=2n$.
The  masses and
three-point couplings are given by the same formulas
as in the $a_{2n-1}^{(2)}$ theory, eqs.(\ref{2.2},\ref{2.3}), but the
fact that
the Coxeter number is even leads to some significant differences.
One-loop mass corrections turn out to have the same form as
for the $a_{2n-1}^{(2)}$ eq.(\ref{2.4}), but opposite signs.

The S-matrix bootstrap starts as in the $a^{(2)}_{2n-1}$ theory
with the  general Ansatz eq.(\ref{Snn}) for $S_{nn}$.
One goes through the
bootstrap to obtain $S_{an}$ and requires that it has a minimal
number of poles. There are some changes in the procedure because the
Coxeter number is even rather than odd. Again one is left
with two possibilities. The one which leads to tree-level
agreement with the amplitude obtained from the lagrangian gives
\EQ
S_{nn}= \prod_{a=1}^{n-1} \frac{(2a)_H(H-2a)_H}{(2a-B)_H(H-2a+B)_H}
\frac{-1}{(\frac{B}{2})_H(H-\frac{B}{2})_H}
\EN
with $H=h-\frac{B}{2}$. The minus sign in this relation between  $H$ and
$B$ is
consistent with the sign of the
one-loop mass shifts.

As compared with the previous situation, we observe that the product
starts at $a=1$ and the two extra blocks have a slightly different
form. An Ansatz which would start at $a=0$ leads to inconsistencies in
the bootstrap, by introducing an additional pole.
The other S-matrix elements $S_{an}$ and $S_{ab}$
have the same form as before.

However there are important differences: as in the $a_{2n-1}^{(2)}$
one must identify in $S_{ab}$ the poles corresponding to the
exchanges of particles $\phi_{(a-b)}$ and $\phi_{(a+b)}$ or
$\phi_{(h-a-b)}$. We are interested in particular in this last
possibility, which occurs for $a+b>\frac{h}{2}$ and, for  the
$a_{2n-1}^{(2)}$ theory, led to the displaced pole situation. The
S-matrix
element is
{\samepage \bea
S_{ab} &=& \prod_{p=1}^b \{2p-b+a-1\}_H\{H-2p+b-a+1\}_H \nonumber\\
&=& {(a-b)_H(a-b+2)_H^{~2}\cdots (a+b-2)_H^{~2}(a+b)_H\over
(a-b+B)_H\cdots\cdots\cdots\cdots(a+b-B)_H}\nopagebreak[3]\\
&&~~~~~~~~\times {(H-a-b)_H(H-a-b+2)_H^{~2}\cdots
(H-a+b-2)_H^{~2}(H-a+b)_H\over(H-a-b+B)_H\cdots\cdots\cdots
\cdots\cdots(H-a+b-B)_H}\nonumber
\ena}
It has a simple pole from a block $(a+b)$ which for $a+b>
\frac{h}{2}$ is displaced
with respect to the pole corresponding to the $s$-channel
exchange of
particle $\phi_{(h-a-b)}$.  This is similar to the situation in the
$a_{2n-1}^{(2)}$ theory but here, since $h$ is even, there is a
nearby double pole from the block $(H-2p+b-a)_H^{~2}$ for $p=\frac{h}{2}
-a$.
In Appendix A we describe how the nature and positions of these
singularities can be explained.
Other singularities can be accounted for in standard fashion. We note
that were it not for the fact
that $H \neq h$, one would encounter quadruple poles, but these
are avoided because of the renormalization of the Coxeter number.

\sect{The $B^{(1)}(0,n)$ theory}

This Lie superalgebra Toda theory may be considered as a truncation of
the two fermion $C^{(2)}(n+1)$ theory which was discussed in
Ref.~\cite{4},
obtained by dropping one of the fermions.
The lagrangian, masses, and bosonic three-point couplings are given by
{\samepage \bea
\b^2{\cal{L}} &=&
- {\frac12} \sum_1^n \phi_a \Box \phi_a
+\frac{i}{2} \bar{\psi} \g \cdot \pa \psi
 \nonumber \\
& -&2\sum_{k=1}^{n-1}  \exp{\left(\sqrt{ \frac{2}{h}}
\sum_{1}^{n-1} m_a \cos{\frac{ak\pi}{n}} ~\phi_a -
\sqrt{\frac{8}{h}}(-1)^k \phi_n \right)}
\nonumber \\
&-&  \exp{\left( \sqrt{\frac{2}{h}} \sum_{1}^{n-1} m_a \phi_a -
\sqrt{\frac{8}{h}}\phi_n \right)}
-  \exp{\left( \sqrt{\frac{2}{h}} \sum_{1}^{n-1} (-1)^a m_a \phi_a -
\sqrt{\frac{8}{h}}(-1)^n \phi_n \right)} \nonumber \\
&-& \frac{1}{\sqrt{2}} \bar{\psi} \psi \exp{\left( \frac{1}{\sqrt{2h}}
\sum_{1}^{n-1} m_a \phi_a - \sqrt{\frac{2}{h}} \phi_n \right)}
\ena}
with
{\samepage \bea
m_{\psi}^2 &=& 2~~~~,~~~
m_a^2 = 8 \sin^2{(\frac{a\pi}{h})}  ~~~~~~~~~~a=1,...,n\nonumber \\
c_{abc} &=& -\frac{\b}{\sqrt{2h}}m_a m_b m_c
=-\frac{4\b}{\sqrt{h}}\D_{abc}
{}~~~~~~~~~{\rm{if}} ~~a\pm b\pm c=0
{}~~~{\rm mod}~h \nonumber \\
c_{abn} &=& \frac{\b}{\sqrt{h}}m_a m_b m_n  = \sqrt{2}
\frac{4\b}{\sqrt{h}}\D_{abn}
{}~~~~~~~~~~~~{\rm{if}} ~~a+b=n
\ena}
and all other bosonic three--point couplings are zero. The Coxeter
number is $h=2n$.
The bosonic part of the lagrangian describes the nonsimply--laced
$c_n^{(1)}$ affine Toda theory which we discuss below.

We begin the bootstrap with the Ansatz
\EQ
S_{\psi \psi}=\prod_{a=0}^{n-1}
\frac{(2a)_H(H-2a)_H}{(2a+B)_H(H-2a-B)_H}
(\frac{B}{2})_H (H-\frac{B}{2})_H
\EN
This expression is the alternative choice one can make when constructing
the S-matrix that led to $S_{nn}$ in the $b_n^{(1)}$ theory, with
$H=h+\frac{B}{2} $ this time. It is not
too difficult to check that it
reproduces the correct tree-level result   which can be extracted from
that given
in eq. (3.4) of Ref.~\cite{4} for the $C^{(2)}(n+1)$ theory,

The S-matrix element $S_{a\psi}$  has the same form as $S_{an}$
in the previous cases and $S_{ab}$ also has the same form
as before. Just as in the $b^{(1)}_n$ theory the analysis of pole and
double pole singularities is affected by the fact that the Coxeter
number is even.

As mentioned above, this theory is obtained from the $C^{(2)}(n+1)$
theory
by dropping one of the fermions.
In Ref.~\cite{6}, we listed separate contributions to the mass
corrections
from the bosons and the fermions of that theory. Therefore, we can
read off the one-loop mass corrections for the present theory:
\EQ
\d\frac{m_a^2}{m_n^2} = - \frac{a\b^2}{4h^2}\sin \frac{2a\pi}{h}
{}~~~~,~~~\d\frac{m_{\psi}^2}{m_n^2} =0
\EN
They are indeed in agreement with the corrections one can read
from our S-matrix at $O(\b^2)$. However, begining at $O(\b^4)$ a new
feature enters: the S-matrix predicts that the mass shifts of
$\phi_n$ and $\psi$ are no longer equal, but in fact $\widetilde{m}_n
< 2\widetilde{m}_{\psi}$. Therefore radiative corrections stabilize the
particle
$\phi_n$, the one-loop threshold effects discussed in Ref.~\cite{13}
are absent at higher order, and the S-matrix bootstrap determines
{\em all} the amplitudes, which was not the case in the $C^{(2)}(n+1)$
theory.

\sect{The $c_n^{(1)}$ theory}

The lagrangian for this theory is obtained by dropping the fermion
of the previous case. The Coxeter number is $h=2n$ and the one loop
mass corrections, from Ref.~\cite{6} are
\EQ
\d\frac{m_a^2}{m_n^2} = - \frac{a\b^2}{2h^2}\sin \frac{2a\pi}{h}
\EN

Here one can start the bootstrap with the Ansatz
\EQ
S_{11} = \frac{(2)_H(H-2)_H}{(2-B)_H(H-2+B)_H} \cdot
\frac{1}{(B)_H(H-B)_H}
\EN
The first fraction accounts for the fact that $c_{11a} \neq 0$ only
if $a=2$ while the second fraction is required for the consistency of
the
bootstrap. Stepping up through the bootstrap one derives then
\EQ\label{6.3}
S_{ab}=\prod_{p=1}^b \{2p+a-b-1\}_H \{H-2p-a+b+1\}_H
\EN
This expression has too many higher order poles unless $H=h \pm B$.
Comparison with the mass corrections gives $H=h+B$.

Our S-matrix is given by the same expression as the bosonic S-matrix for
the
$C^{(2)}(n+1)$ theory of Ref.~\cite{4}, except for the replacement
$h \rightarrow H$.
This replacement has the effect of splitting
the fourth-order poles in eq. (4.26) of that reference into
pairs of double poles.
These separated double poles can be accounted for as in Ref.~\cite{4},
by the dual diagrams in Fig. 9a,b , and Fig. 9c,d respectively,
of that reference.
In addition, if $a+b>n$ one has again a displaced pole situation.
We discuss this in Appendix A.

\sect{The $d^{(2)}_n$ theory}

The lagrangian of this theory is
obtained from the one for the $b_n^{(1)}$ theory by dropping particle
$n$.
The masses and  3-point couplings of the remaining particles
are as in the \tb{} theory.
The Coxeter number is $h=2n$.
The S-matrix is constructed by the same bootstrap procedure as
for the $c_n^{(1)}$ theory and has the form given in \reff{6.3}
but now  $H=h-B$.
We note that compared to the previous case, the change in $H$
eliminates the pole which corresponded to the particle $\phi_n$
of that theory.

\sect{The \tg theory}

Finally we give the S-matrix for the \tg theory.
Its lagrangian is
{\samepage\bea
\b^2 {\cal L} &=& - \frac{1}{2} \phi_1 \Box \phi_1
-\frac{1}{2} \phi_2 \Box \phi_2 -2 \exp(\sqrt{2} \phi_2) \nonumber\\
&&- 3 \exp\left( \frac{1}{\sqrt{6}}\phi_1 -\frac{1}{\sqrt{2}}\phi_2
\right)
-\exp\left( -\frac{\sqrt{3}}{\sqrt{2}}\phi_1 -\frac{1}{\sqrt{2}}\phi_2
\right)
\ena}
and the masses and three point couplings are
\EQ
m_1^2=2 ~~~~,~~~m_2^2 =6,
\EN
\EQ
c_{111}=\sqrt{\frac{8}{3}}\b ~~~~,~~~c_{112}= \sqrt{2}\b
{}~~~~,~~~c_{222}=
3\sqrt{2}\b.
\EN
The Coxeter number is $h=6$.

In the $S_{11}$ amplitude both $\phi_1$ and $\phi_2$ appear as
intermediate states. Because of the particular classical mass ratio,
at the tree level the $s$-channel pole of one particle coincides
with the $u$-channel pole of the other particle.
At the one-loop level the masses receive corrections given by
\EQ
\d \frac{m_2^2}{m_1^2} = \b^2\frac{1}{12\sqrt{3}} \label{gustav}
\EN
which split the above poles. Therefore, starting the bootstrap with
$S_{11}$ presents no problems if the physical masses are
used in the S-matrix.

We begin with a general Ansatz for $S_{11}$ which implements this
pole structure. Because of the $\phi_1$ self-coupling, $S_{11}$ has
to satisfy the bootstrap consistency condition
\EQ
S_{11}(\th ) = S_{11}(\th + \frac{1}{2} \th^1_{11} )
S_{11}(\th - \frac{1}{2} \th^1_{11} )
\EN
with $\th^1_{11} = \frac{2\pi}{3}$. From $S_{11}$ we obtain $S_{12}$ by
bootstrapping on the $\phi_2$ pole. The requirement that $S_{12}$ have
the
minimal number of blocks fixes the expression for $S_{11}$ uniquely and
leads to
\bea
S_{11}(\th )
&=&-\, \frac{\blt{2} \blt{4+\frac{2B}{3}} \blt{H-2}
\blt{H-4-\frac{2B}{3}} }
{\blt{\frac{B}{3}} \blt{2+B} \blt{4+\frac{4B}{3}} \blt{H-\frac{B}{3}}
\blt{H-2-B} \blt{H-4-\frac{4B}{3}} }\nonumber\\
S_{12} (\th )&=& \frac{ \blt{1} \blt{3+\frac{2B}{3}} \blt{H-1}
 \blt{H-3-\frac{2B}{3}} }
{\blt{1+B} \blt{3-\frac{B}{3}} \blt{H-1-B} \blt{H-3+\frac{B}{3}} }\\
S_{22} (\th )&=& -\,\frac{ \blt{2} \blt{2+\frac{B}{3}} \blt{2+
\frac{2B}{3}} }
{\blt{B} \blt{ 2-\frac{B}{3}} \blt{2+B} \blt{2+\frac{4B}{3}} }
\nonumber\\
&&\times
\frac{ \blt{H-2} \blt{H-2-\frac{B}{3}} \blt{H-2- \frac{2B}{3}} }
{\blt{H-B} \blt{ H-2+\frac{B}{3}} \blt{H-2-B} \blt{H-2-\frac{4B}{3}} }
\ena
with $H=6+B$.
We have checked tree-level agreement of these amplitudes. Again
$B=\frac{\b^2}{2\pi} +O(\b^4)$. The corrected masses are again given by
the usual formula $m_a^2 = 8 \sin^2 \frac{\pi a}{H}$, in agreement with
\reff{gustav}.

In the absence of mass corrections $S_{12}$ would have an
anomalous threshold  double pole
at $\th = i \frac{\pi}{2}$ from a box diagram with internal $\phi_1$
lines. This double pole is split into the two simple poles from the
blocks
$\blt{3+\frac{2B}{3}} \blt{H-3-\frac{2B}{3}}$ of the exact S-matrix.
We have checked consistency to $O(\b^4)$. In the $S_{22}$ amplitude,
there is again an interplay between the simple pole due to $\phi_2$
exchange and nearby anomalous threshold singularities.

\sect{Discussion}

We have constructed in this paper elastic S-matrices for all the
families of
nonsimply-laced affine Toda theories. We summarize our results in Table
1.
We have also constructed the S-matrix for \tg. We expect that the
remaining
three cases, namely $d_4^{(3)}$,
$e_6^{(2)}$, and $f_4^{(1)}$ can be constructed in a similar manner.

\def\dash{\mbox{---}}
\renewcommand{\arraystretch}{1.4}
\begin{table}
{\footnotesize\[
\begin{array}{r||c|c|c|c|c|c|}
&\bf
a_{2n-1}^{(2)}&\bf A^{(4)}(0,2n)&\bf
b_n^{(1)}&\bf B^{(1)}(0,n)&\bf c_n^{(1)}&\bf d_n^{(2)}\\ \hline \hline
h&2n-1&2n-1&2n&2n&2n&2n\\ \hline
H&h+{B\over 2}&h-{B\over 2}&h-{B\over2}&h+{B\over 2}&h+B&h-B\\ \hline
m_a^2
&8\sin^2\left({a\pi\over h}\right)
&8\sin^2\left({a\pi\over h}\right)
&8\sin^2\left({a\pi\over h}\right)
&8\sin^2\left({a\pi\over h}\right)
&8\sin^2\left({a\pi\over h}\right)
&8\sin^2\left({a\pi\over h}\right)\\[1mm]\hline
m_n^2&2&\dash&2
&8\sin^2\left({n\pi\over h}\right)
&8\sin^2\left({n\pi\over h}\right)
&\dash\\[1mm] \hline
m_{\psi}^2&\dash&2&\dash&2&\dash&\dash\\
\hline
\epsilon_{ab(a+b)}
&1&1&1&-1&-1&1\\ \hline
\epsilon_{ab(h-a-b)}
&-1&-1&-1&-1&-1&-1\\ \hline
\epsilon_{na(n-a)}
&0&0&0&\sqrt{2}&\sqrt{2}&0\\ \hline
\epsilon_{nna}
&1&0&1&0&0&0\\
\hline\hline
S_{ab}&P_{ab}&P_{ab}&P_{ab}&P_{ab}&P_{ab}&P_{ab}\\ \hline
S_{an}&Q_{a}&\dash&Q_{a}&P_{an}&P_{an}&\dash\\ \hline
S_{nn}&{R_+\over\cbl{B}\cbl{H-B}}&\dash&\frac{R_-}{\cbl{{B\over
2}}\cbl{H-{B\over 2}}}&P_{nn}&P_{nn}&\dash\\[2mm] \hline
S_{a\psi}&\dash&Q_{a}&\dash&Q_{a}&\dash&\dash\\ \hline
S_{n\psi}&\dash&\dash&\dash&Q_{n}&\dash&\dash\\ \hline
S_{\psi\psi}&\dash&R_-
&\dash&R_+{\blt{{B\over 2}}\blt{H-{B\over
2}}\over\blt{B}\blt{H-B}}&\dash&\dash\\[2mm] \hline
\end{array}
\]}
\[\begin{array}{rcl}
c_{ijk}&=&\epsilon_{ijk}\ {4\beta\over\sqrt{h}}\Delta_{ijk}\\[4mm]
P_{ij}&=&\prod_{p=1}^j\cbl{i-j-1+2p}\cbl{H-i+j+1-2p}\\
Q_{i}&=&\prod_{p=1}^i\cbl{{\newh\over 2}-i-1+2p}\qquad
R_\pm=-\prod_{p=1}^{n-1}{(2p)_H(H-2p)_H\over(2p\pm B)_H(H-2p\mp B)_H}
\end{array}\]
\caption{The classical masses, 3-point couplings, and S-matrices of the
nonsimply-laced affine Toda theories.}
\end{table}

The construction was carried out by paralleling that for the
simply-laced
theories, with one important difference: recognizing that radiative
corrections shift the classical masses in a nontrivial manner, we
allowed
coupling-constant dependence in the numerator blocks $\bl{x}$ of the
usual
construction. This led to the possibility of cancellations between
numerator
and denominator blocks and eliminated many unwanted poles that were
produced
in the course of the S-matrix bootstrap. In the end, by insisting on a
minimal
number of poles, we were left with very few possibilities, but
enough to
account for the S-matrices of all the families of nonsimply-laced Lie
algebras
\ta, \tA, \tb, \tB, \tc and \td. There are also a few isolated
possibilities,
which look slightly different, one of which we have shown to describe
the Toda
theory \tg.

These S-matrices do not satisfy all the usual bootstrap equations. In
the
simply-laced theories these equations are a simple consequence of the
identification of all simple poles with elementary particle exchanges.
In the
nonsimply-laced theories however, the simple poles of $S_{ab}$ at
$\th={i\pi\over H}(a+b)$ for $a+b>{h\over 2}$ are not produced just by
the
single particle exchange of $\phi_{h-a-b}$ but involve additional
multiparticle
processes which give rise to anomalous threshold singularities. As a
consequence, the usual bootstrap equations for these poles have to be
replaced
by modified ones which take into account these addtional singularities.

It is remarkable that our S-matrices have a
fairly conventional form in terms of standard blocks, except for the
replacement
of the (integer) Coxeter number $h$ by a renormalized, coupling constant
dependent $H$. In particular, this implies that quantum corrections lead
to
renormalized masses which are given by the usual classical formulas, but
again
with $h$ replaced by $H$.

The same feature
persists for the higher spin conserved charges. We
recall \cite{2} that they are obtained in terms of the Fourier
coefficients of
the matrix
\bea
\varphi_{ab} (\th ) &\equiv& i \frac{d}{d\th} \ln S_{ab}(\th  )
\nonumber\\
&=&\sum_{s=0}^{\infty} \varphi_{ab}^{(s)} e^{-s|\th |}
\ena
Each row and column of this matrix satisfies the charge conservation
equation, as can be seen by inserting the Fourier expansion into the
logarithmic derivative of the S-matrix bootstrap equation:
\EQ
\varphi_{cd}^{(s)} = \varphi_{ad}^{(s)} e^{-s \bar{\th}_{ac}^b}
+\varphi_{bd}^{(s)} e^{s \bar{\th}_{bc}^a}
\EN
(Note however that this relation will be modified in those situations
where
the usual S-matrix bootstrap equation is not valid due to anomalous
threshold
effects, as discussed in \cite{7,8}.)

For the $a_{2n-1}^{(2)}$ and
$b_n^{(1)}$ theories
we find that the even $s$ charges vanish,
while for odd $s$
we obtain for the charges $\g^{(s)}$
\EQ\label{8.6}
\frac{\g_a^{(s)}}{\g_n^{(s)}}=
\frac{\varphi_{ab}^{(s)}}{\varphi_{nb}^{(s)}}
 =(-1)^{{s-1\over2}}2\sin
\frac{\pi sa}{H}
\EN
At the one-loop level this agrees with the explicit values calculated
for the
spin 3 charges of the $a_3^{(2)}$ theory in Refs.~\cite{7,8}. For the
$c_n^{(1)}$
and $d_n^{(2)}$ theories we obtain
\EQ\label{8.7}
\frac{\g_a^{(s)}}{\g_n^{(s)}}={\sin \frac{\pi s a}{H}\over\sin \frac{\pi
sn}{H}}
\EN
and for the \tA and \tB theories
\EQ
\frac{\g_a^{(s)}}{\g_\psi^{(s)}}=
\frac{\varphi_{ab}^{(s)}}{\varphi_{\psi b}^{(s)}}
 =(-1)^{{s-1\over2}}2\sin
\frac{\pi sa}{H}
\EN
For $s=1$ these formulas do of course reproduce the formulas for the
renormalized masses.  Because of the replacement of $h$ by $H$, the
usual fact
that the charges are eigenvalues of the incidence matrix \cite{14} no
longer
holds.

We have checked by comparing the exact S-matrices to one-loop
perturbative
calculations that to $O(\b^4)$ $B=\frac{\b^2}{2\pi}
(1+\frac{\b^2}{4\pi})^{-1}$. For the simply-laced theories, it is well
known
that this particular coupling-constant dependence
implies invariance under the weak-coupling strong-coupling exchange
$\b \rightarrow \frac{4\pi}{\b}$.
This exchange implies $B \rightarrow 2-B$ and leaves invariant the
product
$(x-1+B)(x+1-B)$ which appears in $\{x\}$. Here the situation is
somewhat different because of the additional coupling-constant
dependence in $H$.

Let us consider the $a_{2n-1}^{(2)}$ theory, with the
renormalized Coxeter number $H_{(a)} =2n-1+\frac{B}{2}$. Under the
above substitution
\EQ
H_{(a)} \rightarrow 2n-\frac{B}{2} =H_{(b)}
\EN
where $H_{(b)}$ is the renormalized Coxeter number for the $b_n^{(1)}$
theory.
Furthermore, under the above substitution we have, starting from  the
$a_{2n-1}^{(2)}$ theory
\bea
S_{nn}^{(a)} &=& \prod_{a=0}^{n-1}
\frac{(2a)_H(H-2a)_H}{(2a+B)_H(H-2a-B)_H}
\nonumber\\
&&\rightarrow \prod_{a=0}^{n-1}
\frac{(2a)_H(H-2a)_H}{(2(a+1)-B)_H(H-2(a+1)+B)_H}
\nonumber\\
&=&
- \prod_{a=1}^{n-1}
\frac{(2a)_H(H-2a)_H}{(2a-B)_H(H-2a+B)_H }\cdot \frac{1}{(\frac{B}{2})_H
(H-\frac{B}{2})_H} =S_{nn}^{(b)}
\ena
where for notational simplicity we have omitted the subscripts on the
$H$.
Since all other S-matrix elements are determined by the bootstrap,
this shows that
under the substitution $\b \rightarrow \frac{4\pi}{\b}$
the S-matrices for the $a_{2n-1}^{(2)}$ and $b_n^{(1)}$
theories go into each other. In the same manner one can check
that the S-matrices for  $c_n^{(1)}$ and $d_{n+1}^{(2)}$ go
into each other, as do the S-matrices for the $A^{(4)}(0,2n)$ and
$B^{(1)}(0,n-1)$ theories. In particular, at the symmetric point, $\b^2
=4\pi$ these theories are pairwise equivalent.

We comment on two other features: first, application of the
thermodynamic Bethe Ansatz \cite{14}  leads to the prediction of the
central
charge in the conformal limit, $c=c_{free}$, as one might expect.
Second,
it does not appear that a ``minimal'' S-matrix exists for these
theories.
One cannot drop the blocks involving the dependence on the coupling
constant through the parameter $B$ since these are required for the
consistency
of the bootstrap.

\appendix

\sect{The displaced poles}

In the main text we have drawn the reader's attention to the presence of
simple poles in the exact S-matrix
which are not located precisely at the expected position corresponding
to certain particle masses. As outlined for the case of the
$a_{2n-1}^{(2)}$
theory, the shift in pole positions is due to the
presence of nearby anomalous threshold singularities. Because these
shifted
poles are a novel feature and because we have claimed that they lead to
breakdown of the usual bootstrap procedure, it is important to verify
that all
such poles can be explained by a diagram analysis. In this Appendix we
present
details of this  perturbative verification.

In all the theories that we have considered, the amplitude $S_{ab}$ has
the universal form (choosing $a>b$)
\bea
S_{ab}&=& \prod_p \{2p-b+a-1\}_H\{H-2p+b-a+1\}_H \nonumber\\
&=& \{a-b+1\}_H\{a-b+3\}_H \cdots \{a+b-1\}_H \\
&&\times\{H-a-b+1\}_H \cdots\{H-a+b-1\}_H\nonumber
\ena
It has simple poles from blocks $\blt{a-b}$ and $\blt{H-a-b}$
corresponding
to the
exchange of particle $\phi_{a-b}$ and from blocks
$\blt{a+b}$ and $\blt{H-a-b}$
which in the case $a+b\leq {h\over2}$ correspond to the exchange of
particle
$\phi_{a+b}$. In the case $a+b>{h\over2}$ however, where the couplings
of
the Toda theories show that particle $\phi_{h-a-b}$ is exchanged, these
poles
are not at the expected position.

We study the
poles of $S_{ab}$ in the neighborhood of the point where  the
$s$-channel pole of particle $\phi_{(h-a-b)}$ should appear.  Such
poles come
from the block $\{a+b-1\}_H$. If $h$ is even, the above product also
contains
the blocks $\{H-(h-a-b)-1\}_H\{H-(h-a-b)+1\}_H$ and these too contribute
nearby poles. We will write formulas for this case, with
the understanding that corresponding contributions should be omitted for
those
theories where $h$ is odd. Thus we will consider singularities arising
from
\EQ\label{A.2}
S_{ab} \sim \frac{(a+b)_H}{(a+b-B)_H} \cdot
\frac{[(a+b+H-h)_H]^2}{(a+b+H-h-B)_H
(a+b+H-h+B)_H}
\EN
To the order of our calculation the remaining blocks in
$S_{ab}$ reduce to unity at the poles.

The pole from the block $(a+b)_H$ corresponds to a position in the
$s$-plane
\bea
\widetilde{m}_a^2+\widetilde{m}_b^2+2\widetilde{m}_a\widetilde{m}_b
\cos\frac{\pi}{H}(a+b) &=& 8\sin^2\frac{\pi}{H}(a+b)
\nonumber\\
&=&\widetilde{m}^2_{(h-a-b)}+\d
\ena
where, to lowest order in $\b^2$ (cf. \reff{2.41})
\EQ
\d =8\frac{\pi}{H}(H-h)\D_c ~~~~,~~~
\D_c=-\sin \frac{2\pi}{H}(a+b)
\EN
The pole from $(a+b+H-h)_H$ corresponds to a position
\EQ
\widetilde{m}_a^2+\widetilde{m}_b^2 +2\widetilde{m}_a\widetilde{m}_b
\cos\frac{\pi}{H}(a+b+H-h)
=\widetilde{m}^2_{(h-a-b)} +\hat{\d}
\EN
where
\EQ
\hat{\d} =-4\frac{\pi}{H}(H-h)\hat{\D}_c ~~~~,~~~
\hat{\D}_c = 4\sin\frac{\pi}{H}(a+b)\cos\frac{\pi a}{H}\cos
\frac{\pi b}{H}
\EN
Defining also
\EQ
\s = s-\widetilde{m}^2_{(h-a-b)}
\EN
so that the actual particle pole is at $\s =0$, the poles are located at
$\s =\d$ and $\s = \hat{\d}$ respectively.

The coefficients of the poles are obtained from the general expansions
\bea\label{A.8}
\frac{(x)}{(x+\g )} &=& \frac{1}{\sh (\frac{\th}{2}-\frac{\th_0}{2})}
\left(- \frac{i\pi}{2h}\g +O(\g^2)\right) + regular~terms \nonumber\\
\frac{(x)^2}{(x+\g_1)(x+\g_2)} &=& \frac{1}{\sh
(\frac{\th}{2}-\frac{\th_0}{2})}
 \left( - \frac{i\pi}{2h}(\g_1+\g_2) +O(\g^2)\right) \\
&&+\frac{1}{\sh^2(\frac{\th}{2}-\frac{\th_0}{2})}
\left((\frac{i\pi}{2h})^2 \g_1\g_2 +O(\g^3)\right) +
regular~terms\nonumber
\ena
We will also use $\sh (\frac{\th}{2}-\frac{\th_0}{2} )\sim \frac{1}{2}
(\th -\th_0) \sim -\frac{i}{8 \D} (s-s_0)$ where
\EQ
\D=4 \sin \frac{\pi a}{H} \sin \frac{\pi b}{H}\sin \frac{\pi (a+b)}{H}
\EN
For comparison with perturbation theory it is also convenient to
remove the Jacobian and external particle normalization by multiplying
with the factor $4i m_am_b \sh \th \sim 8i\D$. Finally, to the order
that we are working with, $B\sim \frac{\b^2}{2\pi}$.

We display next the S-matrix poles theory by theory:

\noindent $\bf a_{2n-1}^{(2)}$:
here $H-h = \frac{B}{2}$ and $h=2n-1$ is odd so that the S-matrix has
only
a simple pole
\EQ
S_{ab} \sim \frac{F}{\s -\d}
\EN
with
\EQ\label{a.11}
F=-16i \frac{\b^2}{h}\D^2 ~~~~,~~~\d =2 \frac{\b^2}{h}\D_c
\EN

\noindent $\bf A^{(4)}(0,2n)$:
here $H-h =-\frac{B}{2}$ and $\d = -2\frac{\b^2}{h}\D_c$. Otherwise
the situation is the same as in the previous case.

\noindent $\bf b_n^{(1)}$:
here $H-h = -\frac{B}{2}$ and $h=2n$ is even. We obtain
\EQ
S_{ab} \sim \frac{F}{\s -\d} +\frac{G}{(\s - \hat{\d})^2} +\frac{K}
{(\s -\d )(\s -\hat{\d})^2}
\EN
where
\bea
F&=& -16i \frac{\b^2}{h}\D^2 ~,~~~~
G= -32i \frac{\b^4}{h^2}\D^3 ~,~~~~
K= 64i \frac{\b^6}{h^3}\D^4 ~,\nonumber\\
\d &=& -2\frac{\b^2}{h}\D_c ~,~~~~\hat{\d}=\frac{\b^2}{h}\hat{\D}_c.
\ena

\noindent $\bf B^{(1)}(0,n)$:
this is similar to the $b_n^{(1)}$ case, but with $H-h=\frac{B}{2}$,
$\d =2\frac{\b^2}{h}\D_c$ and $\hat{\d}=-\frac{\b^2}{h}\hat{\D}_c$.

\noindent $\bf  c_n^{(1)}$:
here $H-h=B$ and $h=2n$ is even. We observe that in \reff{A.2} the pole
from the block $(a+b)_H$ cancels against a zero from the denominator
block $(a+b+H-h -B)_H$ so that we have now
\EQ
S_{ab} \sim \frac{F}{\s - \hat{\d}} +
\frac{G}{(\s - \hat{\d})^2}
\EN
with
\beq
F = -16i \frac{\b^2}{h}\D^2 ~,~~~~
G= -64i \frac{\b^4}{h^2}\D^3 ~,~~~~~~
\hat{\d} = -2\frac{\b^2}{h}\hat{\D}_c.
\eeq

\noindent $\bf d_n^{(2)}$:
here $H-h=-B$ so that again the simple pole from $(a+b)_H$ gets
canceled, but also the double pole from $[(a+b+H-h)_H]^2$ gets reduced
to
a simple pole. We have
\EQ
S_{ab} \sim \frac{F}{\s -\hat{\d}}
\EN
with
\beq
F = -16i \frac{\b^2}{h} \D^2 ~,~~~~~~~
\hat{\d} = 2\frac{\b^2}{h} \hat{\D}_c.
\eeq

The pole structure for each theory will be accounted for
by an interplay between single particle poles and anomalous threshold
poles, as we verify case by case.
For the analysis the two dual diagrams in Fig. \ref{dual}
associated with anomalous threshold singularities of
triangle and box diagrams that we encounter below are relevant.
\begin{figure}
\vspace{70mm}
\special{picture fig3 scaled 700}
\caption{Dual diagrams for anomalous threshold poles in
$S_{ab}$.}\label{dual}
\end{figure}
The interested reader should consult Ref.~\cite{4} for an
explanation of the notation and procedure. We list first the geometrical
quantities of interest:

\noindent for Fig. \ref{dual}.a
\bea
\D_a &=& \sin \frac{2\pi}{H}a ~~~~,~~~~\D_b=\sin \frac{2\pi}{H}b
\nonumber\\
\D &=&4 \sin \frac{\pi}{H}a \sin \frac{\pi}{H}b \sin\frac{\pi}{H}(a+b)
\nonumber\\
\D_c &=&\D - \D_a -\D_b =- \sin \frac{2\pi}{H}(a+b)
\ena

\noindent for Fig. \ref{dual}.b
\bea
\hat{\D}_a &=& -4\sin \frac{\pi}{H}a \cos \frac{\pi}{H}b
\cos\frac{\pi}{H}(a+b)
\nonumber\\
\hat{\D}_b &=& -4 \sin \frac{\pi}{H}b \cos\frac{\pi}{H}a
\cos \frac{\pi}{H}(a+b)
\nonumber\\
\D &=&4 \sin \frac{\pi}{H}a \sin \frac{\pi}{H}b \sin\frac{\pi}{H}(a+b)
\nonumber\\
\hat{\D}_c &=&\D - \hat{\D}_a -\hat{\D}_b =4 \sin \frac{2\pi}{H}(a+b)
\cos \frac{2\pi}{H}a \cos\frac{2\pi}{H}b
\ena
All the couplings, pole residues and pole shifts are expressible
in terms of the areas of the various triangles.

We consider now the Feynman diagrams which lead to
poles at the relevant locations.
\begin{figure}
\vspace{60mm}
\special{picture fig4 scaled 850}
\caption{Feynman diagrams leading to the shift of the
$\phi_{h-a-b}$-pole.
Diagrams b),c) and d) contribute only in \ta{} and \tb.}\label{feyn}
\end{figure}
\EQ\label{A.20}
\mbox{Fig. \ref{feyn}.a: }~~~\frac{T_1}{\s} ~~~~~,~~~~
T_1=-ic^2_{ab(h-a-b)}
=-16i\frac{\b^2}{h}\D^2 \EN
This diagram, which corresponds to the actual s-channel exchange of
particle
$\phi_{h-a-b}$, gives the same contribution in all the theories.

The diagrams in Fig. \ref{feyn}.b,c,d exist only in those theories with
$c_{ann}$
couplings, i.e. in $a_{2n-1}^{(2)}$ and $b_n^{(1)}$ and give the
following
contributions:
\EQ\label{A.21}
\mbox{Fig. \ref{feyn}.b: }~~~\frac{T_2}{\s (\s -\d )} ~~~~~,~~~~
T_2=-64i
\frac{\b^4}{h^2}\D^2\D_c
\EN
Here $T_2$ has been computed by including coupling constant factors and
the
residue of the anomalous threshold pole of the triangle diagram derived
in
eq.(4.23) of Ref.~\cite{4}.
We have also included a factor of two for the interchange of the
two vertices.
\EQ\label{A.22}
\mbox{Fig. \ref{feyn}.c: }~~~\frac{T_3}{(\s -\d )^2} ~~~~~,~~~~T_3 = 32i
\frac{\b^4}{h^2}\D^2\D_c
\EN
Here $T_3$ includes coupling constant factors and the coefficient $R_2$
of the
crossed box diagram anomalous threshold double pole, as given in
eq.(4.18) of
Ref.~\cite{4}.
\EQ\label{A.23}
\mbox{Fig. \ref{feyn}.d: }~~~ \frac{T_4}{\s (\s -\d )^2} ~~~~~,~~~~T_4=
-64i
\frac{\b^6}{h^3} \D^2 \D_c^2
\EN
with $T_4$ including coupling constant factors and the square of the
residue of
the triangle anomalous threshold pole.

The diagrams in Fig. \ref{feynh} exist only if $h=2n$, i.e. for the $b$,
$B$,
$c$ and $d$ theories. We have
\begin{figure}
\vspace{60mm}\hspace{5mm}
\special{picture fig5 scaled 850}
\caption{Feynman diagrams leading to the shift of the
$\phi_{h-a-b}$-pole
in theories where $h=2n$.}\label{feynh}
\end{figure}
\EQ
\mbox{Fig. \ref{feynh}.a: }~~~\frac{\hat{T}_2}{\s (\s -\hat{\d})}
{}~~~~~,~~~~\hat{T}_2 = \mp 64i \frac{\b^4}{h^2}\D^2\D_c
\EN
with the plus sign for the $c_n^{(1)}$ theory only;
\EQ
\mbox{Fig. \ref{feynh}.b: }~~~ \frac{\hat{T}_3}{(\s - \hat{\d})^2}
{}~~~~~,~~~~\hat{T}_3 =32i \frac {\b^4}{h^2}\D^2\hat{\D}_c
\EN
\EQ
\mbox{Fig. \ref{feynh}.c: }~~~ \frac{\hat{T}_4}{\s (\s -\hat{\d})^2}
{}~~~~~,~~~~\hat{T}_4=-64i\frac {\b^6}{h^3}\D^2\hat{\D}^2_c
\EN

The diagrams in Fig. \ref{feynt}.a exists only if $h=2n$ and if in
addition the
coupling $c_{an(n-a)}$ exists, i.e. only for the $c_n^{(1)}$ theories.
They give
\begin{figure}
\vspace{60mm}\hspace{5mm}
\special{picture fig6 scaled 850}
\caption{Feynman diagrams leading to the shift of the
$\phi_{h-a-b}$-pole.
Diagrams a) exists in \tc{} and diagram b) in \tb.}\label{feynt}
\end{figure}
\EQ
\mbox{Fig.\ref{feynt}a: }~~~\frac{\tilde{T}_3}{(\s -\hat{\d})^2}
{}~~~~~,~~~~
\tilde{T}_3 = -64i (\D^3 +\D^2\hat{\D}_c)
\EN

The diagram in Fig.\ref{feynt}b
exists only if $h=2n$ and $c_{ann}$ exists, i.e. only for $b_n^{(1)}$.
Including a factor of two for the interchange of the two triangles, we
have
\EQ
\mbox{Fig.\ref{feynt}b: }~~~\frac{\tilde{T}_4}{\s (\s - \d )(\s -
\hat{\d})}
{}~~~~~,~~~~ \tilde{T}_4= -128i \frac{\b^6}{h^3} \D^2\D_c \hat{\D}_c
\EN

We consider now each theory, and compare the pole contributions
from the S-matrix with the corresponding poles of Feynman diagrams.

\noindent $\bf a_{2n-1}^{(2)}$: we want to show that
\EQ
\frac{F}{\s -\d} =\frac{T_1}{\s} +\frac{T_2}{\s (\s - \d )} +
\frac{T_3}{(\s -\d )^2} + \frac{T_4}{\s (\s - \d )^2}
\EN
This requires
\bea
&&T_1 = F \nonumber\\
&&2\d T_1 -T_2 -T_3 =  \d F \nonumber\\
&&\d^2T_1 -\d T_2 +T_4 =0
\ena
Using the expressions in
eqs.(\ref{a.11},\ref{A.20},\ref{A.21},\ref{A.22} and
\ref{A.23}) one can check that these conditions are indeed satisfied.

\noindent $\bf b_n^{(1)} $:
we have to show that
\bea
\frac{F}{\s -\d} +\frac{G}{(\s - \hat{\d})^2} +\frac{K}{(\s -\d )
(\s - \hat{\d})^2} =\frac{T_1}{\s}+\frac{T_2}{\s (\s - \d)} +\frac
{T_3}{(\s - \d )^2} \nonumber\\
+\frac{T_4}{\s (\s - \d )^2} +\frac{\hat{T}_2}{\s (\s - \hat{\d})}
+\frac{\hat{T}_3}{(\s - \hat{\d})^2} +\frac{\hat{T_4}}{\s (\s -
\hat{\d})^2} +\frac{\tilde{T}_4}{\s (\s - \d )(\s - \hat{\d})}
\ena
from which we derive the following conditions:
\bea
a)&&T_1=F \nonumber\\
b)&&\d T_1-T_2-\hat{T}_2-T_3-\hat{T}_3 =-G \nonumber\\
c)&&(-\d^2+2\d \hat{\d})T_1 +(\d -2 \hat{\d})T_2 -\hat{\d}\hat{T}_2
+(-2\hat{\d} +2\d )T_3 \nonumber\\
&&~~~~+T_4+\hat{T}_4+\tilde{T}_4 =K \nonumber\\
d)&&(-\d \hat{\d}^2-2\d^2\hat{\d}+\d^3)T_1 +(\hat{\d}^2+2\d
\hat{\d}-\d^2)T_2
+2\d \hat{\d}\hat{T}_2 \nonumber\\
&&~~~~+(\hat{\d}^2-\d^2)T_3
-2\hat{\d}T_4 -2\d\hat{T}_4 -(\d +\hat{\d})\tilde{T}_4 = -\d K
\nonumber\\
e)&&\d^2\tilde{\d}^2T_1 - \d \hat{\d}^2T_2 -\d^2\hat{\d}\hat{T}_2
+\hat{\d}^2T_4
+\d^2\hat{T}_4 +\d \hat{\d}\tilde{T}_4=0
\ena
With the expressions given earlier, it is not too difficult to check
that the conditions a), b) and e) are indeed satisfied, as a
result
of the interplay between "hatted" and "unhatted" diagrams using also the
identity $2\D_c+\hat{\D}_c =\D$, but that the
conditions c) and d) are not. However, there are additional
two-loop
diagrams that must be considered in this case: the one in Fig.
\ref{extra}.a
and the one obtained by interchanging top and bottom rungs.
\begin{figure}
\vspace{85mm}
\special{picture fig7 scaled 850}
\caption{An additional diagram contributing in the \tb theory.
a)  Feynman diagram, b) Dual diagram.}\label{extra}
\end{figure}
The corresponding dual diagram in Fig. \ref{extra}.b would be planar
and therefore lead to a triple pole if the classical masses were used.
If one
uses the actual masses the diagram is not quite planar, but those dual
diagrams
associated with reduced diagrams are, leading to a splitting of the
triple pole
into a simple and a double pole. Although we have not checked the
details, it is
plausible that one obtains a contribution
\EQ
\frac{T_5}{(\s - \d )^2(\s - \hat{\d})} +\frac{\hat{T}_5}{(\s - \d )
(\s - \hat{\d})^2}
\EN
with
\EQ
T_5 = 64i \frac{\b^6}{h^3}\D^3\D_c ~~~~,~~~~\hat{T}_5 = 64i
\frac{\b^6}{h^3}\D^3(\D_c+\hat{\D}_c).
\EN
Including this contribution does not affect the three conditions that
were satisfied previously, but leads now to satisfaction of the other
two,
and therefore an explanation of the pole structure for the $b_n^{(1)}$
theory.

\noindent $\bf c_n^{(1)}$:
we must show that
\EQ
\frac{F}{(\s - \hat{\d})}+\frac{G}{(\s - \hat{\d})^2} = \frac{T_1}{\s}
+\frac{\hat{T}_2}{\s (\s - \hat{\d})} +\frac{\hat{T}_3}{(\s -
\hat{\d})^2}
+\frac{\hat{T}_4}{\s (\s - \hat{\d})^2} +\frac{\tilde{T}_3}{(\s -\hat
{\d})^2}
\EN
leading to the conditions
\bea
&&T_1=F \nonumber\\
&&2\hat{\d}T_1-\hat{T}_2-\hat{T}_3-\tilde{T}_3=\hat{\d}F-G \nonumber\\
&&\hat{\d}^2T_1-\hat{\d}\hat{T}_2 +\hat{T}_4 =0
\ena
which are indeed satisfied.

\noindent $\bf d_n^{(2)}$:
we must show
\EQ
\frac{F}{\s - \hat{\d}} = \frac{T_1}{\s} +\frac{\hat{T}_2}{\s (\s -
\hat{\d})}
+\frac{\hat{T}_3}{(\s - \hat{\d})^2} +\frac{\hat{T}_4}{\s (\s -
\hat{\d})^2}
\EN
leading to
\bea
&&T_1=F \nonumber\\
&&\hat{\d}T_1 -\hat{T}_2 -\hat{T}_3=0 \nonumber\\
&&\hat{\d}^2T_1 -\hat{\d}\hat{T}_2 +\hat{T}_4=0
\ena
which are again satisfied.

This completes the perturbative verification of the position and
residues of the displaced poles in these bosonic exact S-matrices. We
have
not
examined in detail the corresponding situation for the theories with
fermions or for \tg.

\sect{The coupling constant dependence}

For the simply-laced theories it has been checked in perturbation theory
\cite{11} that the coupling constant dependence of
the S-matrix arises through the
combination $B= \frac{1}{2\pi}\frac{\b^2}{1+\frac{\b^2}{4\pi}}$. In
this appendix we carry out a similar perturbative verification, at the
one-loop level, for two
nonsimply-laced cases. We compute from the lagrangian,
to $O(\b^4)$, the residue of  a suitable S-matrix
element at a single-particle pole, and compare to the same expression
obtained
from the exact S-matrix.
To make the calculations manageable, we restrict ourselves
to cases with only two fields.

We consider first
the $a^{(2)}_3$ theory, with fields $\phi_1$ and $\phi_2$, classical
masses
$m_1^2=6$, $m_2^2=2$, and relevant couplings
\EQ
{\cal L}_{int}= -\phi_1^3 +\phi_1\phi_2^2 -\frac{3}{4}\phi_1^4
-\frac{1}{2}\phi_1^2\phi_2^2 -\frac{1}{12}\phi_2^4 +O(\phi^5)
\EN
We will compute the residue of the S-matrix element $S_{22}$ at the pole
corresponding to exchange of the particle $\phi_1$. We find it
convenient to
work with renormalized masses, changing however the overall mass scale
so as to
keep the mass of particle $\phi_2$ at its classical value $m_2^2=2$.
This leads
to an overall rescaling of the coupling constants since the mass scale
enters in
their definition. Additional $O(\b^4)$ contributions arise from one-loop
vertex
corrections and wave-function renormalization.

The one-loop corrections to
the $\langle 1,2,2 \rangle$ vertex
are represented  by the diagrams in Fig. \ref{8}.a,b
and give, respectively from the bubbles and the triangles,
\begin{figure}
\vspace{115mm}
\special{picture fig8 scaled 850}
\caption{Feynman diagrams giving the one-loop correction to the $\langle
1,2,2
\rangle$ vertex in the $a_3^{(2)}$ theory}\label{8}
\end{figure}
\EQ
(a): -\frac{i\b^3}{2\sqrt{3}} ~~~~,~~~~(b):
i\b^3\left(-\frac{1}{36\sqrt{3}}+
\frac{1}{6\pi}\right)
\EN
to be compared with the classical contribution $\langle 1|i{\cal
L}|2,2\rangle=2i\b$.

Additionally, there are contributions from
wave-function renormalization,
represented in Fig. \ref{8}.c, which give
\EQ
Z_1^{\frac{1}{2}}Z_2= 1+{\b^2\over 2}\left( \frac{5}{36\sqrt{3}}
-\frac{1}{2\pi}\right)
\EN
to be multiplied into the classical coupling.
Finally, there is an additional contribution to the coupling
constant from the mass rescaling we performed in order to
keep the mass of the particle $\phi_2$ at its classical
value. This amounts to multiplying the classical coupling constant by
\EQ
1-\frac{\d m_2^2}{m_2^2} =1+\frac{\b^2}{12\sqrt{3}}
\EN
obtained from Ref.~\cite{6}.
The total $O(\b^3)$ value of the coupling is thus
\EQ
\langle 1,2,2\rangle= 2i\b
-2i\b^3\left(\frac{1}{9\sqrt{3}}+\frac{1}{6\pi}\right)
\EN
and it leads to a contribution to the $S_{22}$ amplitude, in the
vicinity of the $\phi_1$ particle pole
\EQ
\frac{1}{4\mt_2^2\sh \th}
\left(2i\b
-2i\b^3\left(\frac{1}{9\sqrt{3}}+\frac{1}{6\pi}\right)\right)^2
\frac{1}{s-\widetilde{m}_1^2}
\EN
to be compared to order $\b^4$ with the corresponding pole from the
exact
S-matrix
\EQ
S_{22}=\frac{(2)_H(H-2)_H}{(B)_H(2+B)_H(H-2-B)_H(H-B)_H}
\EN
We find
\EQ
B= \frac{\b^2}{2\pi} -\frac{\b^4}{8\pi^2}
\EN
which is indeed  consistent with the expected form of $B$.

We have carried out the same calculation for the $c_2^{(1)}$ theory by
studying the behavior of the $S_{12}$ amplitude in the vicinity of the
$\phi_1$ pole. Here the classical masses are $m_1^2=4$ and $m_2^2=8$,
and the
relevant part of the interaction lagrangian is
\EQ
{\cal L}_{int}= 2\sqrt{2}\phi_1^2\phi_2 -\frac{1}{3}\phi_1^4
-2\phi_1^2\phi_2^2
-{2\over 3}\phi_2^4+O(\phi^5).
\EN
\begin{figure}
\vspace{50mm}
\special{picture fig9 scaled 780}
\caption{Feynman diagrams giving the one-loop correction to the
$\langle 1,1,2 \rangle$ vertex in the $c_2^{(1)}$ theory}\label{9}
\end{figure}
The one loop vertex corrections from Fig. \ref{9}.a,b are
\EQ
(a): -\frac{3i\b^3}{\sqrt{2}}
{}~~~~,~~~~(b):i\b^3\left(\frac{1}{2\sqrt{2}}
+\frac{1}{\sqrt{2}\pi}\right)
\EN
as compared to the tree level value $\langle 1,1|i{\cal L}|2\rangle=
4\sqrt{2}i\b$. The wave-function renormalization factors represented in
Fig.
\ref{9} give
\EQ
Z_1Z_2^{\frac{1}{2}} = 1+\b^2\left(\frac{1}{16} -\frac{5}{16\pi}\right)
\EN
and from the shift in the mass scale one obtains an additional
correction to the vertex
\EQ
1-\frac{\d m_2^2}{m_2^2} =1+\frac{\b^2}{16}
\EN
leading to a total value of the coupling
\EQ
\langle 1,1,2\rangle=4\sqrt{2}i\b -i\b^3
\frac{3\sqrt{2}}{4}\left(1+\frac{1}{\pi}\right)
\EN
Comparison to the exact S-matrix $S_{12}=\{2\}_H\{H-2\}_H $
again checks the correct identification of $B$.

\end{document}